\documentclass[a4paper,11pt]{article}
\usepackage{jheppub}
\usepackage{lineno}
\usepackage{microtype}
\usepackage[T1]{fontenc}
\usepackage{eurosym}
\usepackage{booktabs}
\usepackage{multirow}
\usepackage{graphicx}
\usepackage{amsmath}
\usepackage{colortbl}
\usepackage{amssymb}
\usepackage{todonotes}
\usepackage{wrapfig}
\definecolor{DeepPink}{rgb}{0.796,0.004,0.384}
\definecolor{RoyalBlue}{rgb}{0.02,0.016,0.667}
\definecolor{LavenderBlush}{rgb}{1.00,0.941,0.961}
\usepackage{lastpage}
\usepackage{enumitem}
\usepackage{lipsum}
\usepackage{physics}
\usepackage{longtable}
\usepackage[toc]{multitoc}
\usepackage{comment}
\usepackage{array}

\numberwithin{equation}{section}

\newcommand{\ssfill}{\xleaders\hbox to 0.35em{\scriptsize.}\hfill}
\makeatletter
\newcommand{\oset}[3][0ex]{\mathrel{\mathop{#3}\limits^{\vbox to#1{\kern-2\ex@\hbox{$\scriptstyle#2$}\vss}}}}
\makeatother
\newcommand*{\cventry}[7][.25em]{\noindent\begin{tabular*}{\textwidth}{l@{\extracolsep{\fill}}r}{\bfseries #4} & {\bfseries #5} \\{\itshape #3\ifthenelse{\equal{#6}{}}{}{, #6}} & {\itshape #2}\\\end{tabular*}\ifx&#7&\else{\\\vbox{\small#7}}\fi\par\addvspace{#1}}
\newcommand*{\hintfont}{\bfseries}
\newcommand*{\hintstyle}[1]{{\noindent\hintfont{#1}}}
\newcommand*{\cvitem}[3][.25em]{\ifthenelse{\equal{#2}{}}{}{\hintstyle{#2}: }{#3}\par\addvspace{#1}}
\let\OriginalQuotation\quotation
\renewcommand*{\quotation}{\OriginalQuotation\small\sf}

\newcommand{\be}{\begin{equation}}
\newcommand{\ee}{\end{equation}}
\newcommand{\ba}{\begin{array}}
\newcommand{\ea}{\end{array}}
\newcommand{\baa}{\begin{array}}
\newcommand{\eaa}{\end{array}}
\newcommand{\bea}{\begin{eqnarray}}
\newcommand{\eea}{\end{eqnarray}}

\newcommand{\MS}{{\overline{\rm MS}}}
\newcommand{\LO}{{\rm LO}}
\newcommand{\NLO}{{\rm NLO}}
\newcommand{\LOp}{{\rm LO+}}
\newcommand{\NLOp}{{\rm NLO+}}

\newcommand{\LMS}{\Lambda_{\overline{\rm 	MS}}}
\newcommand{\vv}{{\langle \mathbf{v}^2 \rangle}}
\newcommand{\pp}{{\langle \mathbf{p}^2 \rangle}}

\newcommand\colorrow{\rowcolor{LavenderBlush}}

\newcolumntype{\Vert}{!{\vrule width 1pt}}
  
\arxivnumber{2506.11958}

\title{\boldmath Temperature Dependence of Heavy Quark Diffusion from (2+1)-flavor Lattice QCD}

\author[a]{Dennis Bollweg}
\author[b]{Jorge Luis Dasilva Gol\'an}
\author[c]{Olaf Kaczmarek}
\author[c]{Rasmus Norman Larsen}
\author[d]{Guy D. Moore}
\author[b]{Swagato Mukherjee}
\author[b]{Peter Petreczky}
\author[e]{Hai-Tao Shu}
\author[d]{Simon Stendebach}
\author[d,f]{Johannes Heinrich Weber}

\affiliation[a]{Computing and Data Sciences Directorate,Brookhaven National Laboratory, Upton, New York 11973, USA}
\affiliation[b]{Physics Department, Brookhaven National Laboratory, Upton, New York 11973, USA}
\affiliation[c]{Fakult\"at f\"ur Physik, Universit\"at Bielefeld, D-33615 Bielefeld, Germany}
\affiliation[d]{Institut f\"ur Kernphysik, Technische Universit\"at Darmstadt, Schlossgartenstra\ss e 2, D-64289 Darmstadt, Germany}
\affiliation[e]{Key Laboratory of Quark \& Lepton Physics (MOE) and Institute of Particle Physics, Central China Normal University, Wuhan 430079, China}
\affiliation[f]{Institut f\"ur Physik \& IRIS Adlershof, Humboldt-Universit\"at zu Berlin, D-12489 Berlin, Germany}

\emailAdd{dbollweg@bnl.gov}
\emailAdd{jgolandas@bnl.gov}
\emailAdd{okacz@physik.uni-bielefeld.de}
\emailAdd{rasmusnormannlarsen@gmail.com}
\emailAdd{guy.moore@physik.tu-darmstadt.de}
\emailAdd{swagato@bnl.gov}
\emailAdd{petreczk@bnl.gov}
\emailAdd{hai-tao.shu@ccnu.edu.cn}
\emailAdd{simon.stendebach@tu-darmstadt.de}
\emailAdd{dr.rer.nat.weber@gmail.com}

\abstract{
We present a lattice determination  of the heavy-quark diffusion coefficient in (2+1)-flavor QCD with almost physical quark masses. The momentum and spatial diffusion coefficients are extracted for a wide temperature range, from $T=163$ MeV to $10$ GeV.  The results are in agreement with previous works from the HotQCD collaboration, and show fast thermalization of the heavy quark inside the QGP. Near the chiral crossover temperature  $T_c\simeq150$ MeV, our results are close to the AdS/CFT estimation computed at strong coupling. 
}

\begin{document}

\begin{flushright}
  HU-EP-25/21-RTG
\end{flushright}

\maketitle
\flushbottom

\section{Introduction}
\label{sec:introduction}

The study of heavy quarks in the quark-gluon plasma (QGP) \cite{PhysRevLett.34.1353,STAR:2005gfr,BRAHMS:2004adc} is essential for understanding the general properties of strongly interacting matter produced in heavy-ion collisions (HIC). Heavy quarks, such as charm and bottom quarks, provide unique insights into the non-equilibrium dynamics of the QGP, as they are produced in the early stages of the collision and then propagate throughout the evolving plasma. Therefore, the study of interaction and equilibration of heavy quarks is crucial for understanding many properties of the QGP, such as its near-perfect fluid behavior \cite{Rapp:2018qla,Dong:2019byy,He:2022ywp}. 

The interaction of heavy quarks with the medium can be effectively described by Langevin dynamics \cite{Moore:2004tg} for sufficiently large masses, and their motion is characterized with two transport coefficients: the heavy-quark momentum diffusion coefficient $\kappa$ and the drag coefficient $\eta$. These transport coefficients play a central role, since they determine the momentum distributions of the heavy-flavor hadrons in the HIC, which have been precisely measured at RHIC \cite{PHENIX:2006iih,STAR:2006btx} and LHC \cite{ALICE:2012ab}. 

The momentum diffusion coefficient $\kappa$ quantifies the average momentum transferred from the QGP to a heavy quark by random, uncorrelated in time, momentum kicks, with $3\kappa$ being the mean squared momentum transfer per unit time. Physically, it describes how much momentum a heavy quark loses and how fast it thermalizes in the plasma. For sufficiently heavy quarks, the relaxation time is expected to be $M/T$ times larger than that of light quarks \cite{Moore:2004tg,PhysRevD.37.2484}, where $M$ is the mass of the heavy quark and $T$ is the temperature of the medium. Recent measurements have confirmed that this relaxation time is indeed shorter than the mean lifetime of the QGP (even with the enhancement of the prefactor $M/T$), suggesting a rather fast relaxation of heavy quarks and confirming the strongly interacting nature of the matter produced in HIC. Therefore, because of its relevance to the phenomenology of heavy ions, an accurate estimate of $\kappa$ has become a critical goal for both theoretical and experimental studies. 

Lattice field theory techniques provide a first-principles approach to the non-perturbative aspects of quantum chromodynamics (QCD), including the extraction of $\kappa$. However, the calculation is far from trivial, since it requires a spectral reconstruction of the correlation functions computed numerically on the lattice. Like all transport coefficients, the diffusion coefficient is directly encoded in the low-frequency behavior of the spectral function. It is precisely this inversion problem that makes the extraction extremely challenging, i.e., even when the correlators are determined precisely, the spectral functions extracted from them —and consequently the diffusion coefficient— can still carry large uncertainties. 

The heavy-quark diffusion coefficient has been calculated in perturbation theory \cite{PhysRevD.37.2484,Moore:2004tg,Caron-Huot:2007rwy}, but the calculation is reliable only for asymptotically large temperatures $T$. In the present work, we build on previous studies to extract the heavy-quark diffusion coefficient from lattice QCD \cite{Banerjee:2011ra,Francis:2015daa,Altenkort:2020fgs,Banerjee:2022gen,Brambilla:2022xbd,Brambilla:2020siz,Altenkort:2023oms,Altenkort:2023eav,Banerjee:2022uge}. The HotQCD collaboration calculated $\kappa$ in a (2+1)-flavor lattice QCD framework for physical strange quarks and rather unphysical light quarks with mass $m_l=m_s/5$ \cite{Altenkort:2023eav,Altenkort:2023oms} corresponding to the pion mass of  $m_\pi\simeq320$ MeV, and for a temperature range between $T=195$ and $352$ MeV. The effect of light-quark masses is expected to be irrelevant for sufficiently large temperatures, but important for temperatures below $250$ MeV. Thus, the main goal of this work is to extend the previous calculation to an almost physical point, with light-quark masses of $m_l=m_s/20$ corresponding to a pion mass of $m_\pi\simeq160$ MeV, and for a wider temperature range, from values close to the crossover temperature $T_c$ up to the GeV scale. The extension to high temperature is needed for making contact with perturbation theory.

The paper is organized as follows. In Sec.~\ref{sec:theoretical-framework} we present the theoretical framework for the extraction of the diffusion coefficient from the first-principles lattice calculation. In Sec.~\ref{subsec:lattice} we give the details of the lattice simulations, such as the simulation parameters for the data ensembles and our choice of renormalization/regularization method. In Sec.~\ref{subsec:norms} we discussed our normalization choices, and in Secs.~\ref{subsec:continuum} and~\ref{subsec:flow} we describe very sophisticated techniques for continuum and zero-flow extrapolation of the correlation functions already used in previous works, with special emphasis on the ultraviolet (UV) regularization of both the electric and magnetic components of the correlators. Sec.~\ref{sec:perturbative_results} provides an extensive overview of literature results on the diffusion coefficient and spectral functions in perturbation theory and the connection between the perturbative results and lattice QCD results at the level of the correlation functions. We introduce the spectral reconstruction method in Secs.~\ref{subsec:spectral_E} and~\ref{subsec:spectral_B}, where we use the leading (LO) and next-to-leading (NLO) order perturbative QCD results to model the UV behavior of the spectral function, while leaving the infrared (IR) behavior completely determined by the diffusion coefficient itself. A key aspect of this section is the fitting strategy used to interpolate between the IR and UV regimes of the spectral function. We explore different theoretically motivated functional forms to describe the transition between these regimes, allowing us to constrain the value of $\kappa$ by taking into account as much model dependence as possible. We present the final result for the momentum and spatial diffusion coefficient in Sec.~\ref{subsubsec:kappa_D}. Finally, the paper ends in Sec.~\ref{sec:conclusions} with some conclusions and outlooks. 


\section{Theoretical framework}
\label{sec:theoretical-framework}

Close to equilibrium, the response of strongly interacting matter to small perturbations can be studied in terms of a handful of transport coefficients, such as the charge diffusion coefficient, or the shear and bulk viscosities. These coefficients can be extracted from current-current correlation functions, where the type of current determines the specific coefficient to be computed. Among all the coefficients, the easiest one to extract is the heavy-quark momentum diffusion coefficient $\kappa$, which is formally defined as the mean-squared momentum transfer per unit time experienced by a heavy quark immersed in a QGP. Together with the equilibrium mean-squared momentum that the heavy quark carries in the medium, this determines how long it takes for the quark to thermalize in the QGP. For a more extended review see section 6 of \cite{Ding:2015ona}.

The Green-Kubo relation characterizes the linear response of the thermalized QGP to small perturbations induced by the motion of particles. According to it, the heavy-quark diffusion coefficient can be extracted from the spectral reconstruction of the vector-vector correlation function (at thermodynamic equilibrium) in the following way:
\be
	G^{ii}(\tau,\vb{p},T)=\int \mathrm{d}\vb{x}\exp{i\vb{x}\vb{p}}\langle J^i_V(0,0,T)J^i_V(\tau,\vb{x},T)\rangle=\int_0^\infty \frac{\mathrm{d}\omega}{\pi}\rho^{ii}(\omega,\vb{p},T)K(\omega,\tau,T)\,,
    \label{eq:current-current-correlation}
\ee
using the corresponding vector current $J^\mu_V(t,\vb{x},T)=\bar{\psi}(t,\vb{x},T)\gamma^{\mu}\psi(t,\vb{x},T)$, and the integration Kernel
\be
    K(\omega,\tau,T)=\frac{\cosh(\omega\tau-\omega/2T)}{\sinh(\omega/2T)}\,.
	\label{eq:kernel}
\ee
The spatial heavy-quark diffusion coefficient is then defined as the zero-frequency and zero-momentum limit of the spectral function:
\be
    D_s(T)=\frac{1}{3\chi_q}\lim_{\vb{p}\rightarrow 0}\lim_{\omega\rightarrow 0}\frac{\rho^{ii}(\omega,\vb{p},T)}{\omega}\,,
	\label{eq:zero-limit}
\ee
where $\chi_q$ is the heavy quark number susceptibility, $q=c,b$ for charm and bottom, respectively. The determination of $D_s$ on the lattice through Eq.~\eqref{eq:zero-limit} is extremely challenging, as discussed in Ref.~\cite{Petreczky:2008px}. An alternative approach relies on the use of the heavy-quark effective theory (HQET) \cite{Eichten:1990vp} as discussed in Refs.~\cite{Caron-Huot:2009ncn,Bouttefeux:2020ycy,Casalderrey-Solana:2006fio}. In the HQET approach, the current-current correlation function is expressed in terms of the chromo-electric and chromo-magnetic correlation functions
\cite{Caron-Huot:2009ncn,Bouttefeux:2020ycy}:
\begin{eqnarray}
	G_E(\tau,T)=-\sum_{i=1}^3\frac{\langle\Re\Tr(U(\beta,\tau)E_i(\tau,\vb{0})U(\tau,0)E_i(0,\vb{0}))\rangle}{3\langle\Re\Tr U(\beta,0)\rangle}\,,
	\label{eq:corr_E}\\
	G_B(\tau,T)=\sum_{i=1}^3\frac{\langle\Re\Tr(U(\beta,\tau)B_i(\tau,\vb{x})U(\tau,0)B_i(0,\vb{x}))\rangle}{3\langle\Re\Tr U(\beta,0)\rangle}\,,
	\label{eq:corr_B}
\end{eqnarray}
where $\beta=1/T$ is the inverse temperature, $\tau$ is the Euclidean time separation, and $U(\tau_1,\tau_2)$ is a temporal Wilson line between $\tau_1$ and $\tau_2$. In these expressions, the correlation functions have been normalized by the mean trace of the Polyakov loop $\mathcal{P}=\langle\Re\Tr U(\beta,0)\rangle$. The correlators defined above have the following spectral representation \cite{Caron-Huot:2009ncn,Bouttefeux:2020ycy}:
\begin{equation}
    G_{E,B}(\tau,T)=\int_0^{\infty} \frac{\mathrm{d} \omega}{\pi} \rho_{E,B}(\omega,T) K(\omega,\tau,T)\,.
    \label{eq:spectral_rep}
\end{equation}
The $\omega\rightarrow 0$ limit of the corresponding spectral functions defines the transport coefficients $\kappa_E$ and $\kappa_B$ \cite{Caron-Huot:2009ncn,Bouttefeux:2020ycy}:
\begin{equation}
    \kappa_{E,B}=2 T \lim_{\omega \rightarrow 0} \rho_{E,B}(\omega,T)/\omega\,.
    \label{eq:kappaE_B}
\end{equation}
Thus, the heavy-quark momentum diffusion coefficient, up to corrections ${\cal O}(T^2/M^2)$, can be written as \cite{Bouttefeux:2020ycy}:
\begin{eqnarray}
    \kappa=\kappa_E+\frac{2}{3} \langle \mathbf{v^2} \rangle \kappa_B\,.
\end{eqnarray}
The second term in the above equation describes the leading mass dependence of the coefficient. This dependence arises because an equilibrated, non-infinitely heavy quark carries a mean-squared velocity in the medium and therefore, interacts with color-magnetic as well as color-electric fields. For large masses, we expect $\vv \simeq 3T/M$. Then, the momentum-space diffusion coefficient $\kappa$ is related to the spatial heavy-quark diffusion coefficient $D_s$ via an Einstein relation:
\be
	D_s=\frac{2T^2}{\kappa}\frac{\pp}{3MT}\,,
	\label{eq:D_s}
\ee
where $\pp$ is the average thermal momentum of the heavy quark.

The chromo-electric and chromo-magnetic correlation functions are constructed in terms of gluonic operators, which suffer from strong short-distance fluctuations that yield low signal-to-noise ratio. Accurate estimation of the correlation functions \eqref{eq:corr_E} and \eqref{eq:corr_B} thus requires noise reduction techniques, such as multi-level updating \cite{Luscher:2001up}. While the electric correlation function requires only a finite renormalization at finite lattice spacing, and is regulator independent within the $\MS$ scheme \cite{Christensen:2016wdo}, the magnetic correlator involves a non-trivial anomalous dimension \cite{Eichten:1990vp}, and therefore it requires a specific renormalization method. On the lattice, we will use the gradient flow (GF) renormalization procedure \cite{Narayanan:2006rf,Luscher:2009eq,Luscher:2010iy,Luscher:2011bx} to solve both problems at once, which has been shown to be as good as multi-level algorithms for noise reduction in the quenched theory \cite{Altenkort:2020fgs,Brambilla:2022xbd}, while providing renormalized field operators for any flow time $\tau_F$ larger than the square of the lattice-spacing scale. We will explain our choice of renormalization method as well as a small description of our GF implementation in Sec.~\ref{sec:numerical-results}. 

In both, the electric and the magnetic case, spectral reconstruction is a crucial and difficult aspect of the calculation. The main problem arises from the fact that the correlation functions are computed only at some discrete Euclidean time intervals $\tau$. Thus, to extract the spectral function $\rho(\omega,T)$ from these data, one has to use a theoretically motivated Ansatz. For this purpose we can use the perturbative QCD (pQCD) estimates of the spectral function. At LO and NLO \cite{Caron-Huot:2009ncn}, the spectral function in pQCD is predicted to behave as $\rho(\omega,T) \propto \omega^3$ for large $\omega$, while the low-energy behavior of the spectral function directly yields the diffusion coefficient $\kappa$ as $\rho(\omega,T) \propto \kappa_{E,B}\omega$. To interpolate between these two regions, we will explore several methods, like for instance sharp, smooth, or polynomial transition between the IR and UV behavior of the spectral function. These methods are discussed in detail in Sec.~\ref{sec:spectral}. In order to determine $\kappa$ and $D_s$, we also need to determine the average thermal velocity, $\vv$ and averaged thermal momentum, $\pp$ of the heavy quark. Following previous works \cite{Altenkort:2023eav,Altenkort:2023oms}, in Sec.~\ref{subsubsec:QPM} we will determine these quantities in terms of a quasi-particle model (QPM).


\section{Lattice QCD results on the chromo-electric and chromo-magnetic correlators}
\label{sec:numerical-results}

\subsection{Lattice Setup}
\label{subsec:lattice}

\begin{table}[ht]
\centering
\begin{tabular}{ccccccccc}
\toprule
$T$ [MeV] & $\beta$ & $a$ [fm] & $a \cross m_s$ & $a \cross m_l$ & $N_{\sigma}$ & $N_{\tau}$ &  \# conf. & streams \\
\midrule
149	&	7.3730	&	0.0602	&	0.02500	&	0.00125	&	64	&	22	&	4663		&	35	\\
164	&	7.3730	&	0.0602	&	0.02500	&	0.00125	&	64	&	20	&	7424		&	36	\\
182	&	7.3730	&	0.0602	&	0.02500	&	0.00125	&	64	&	18	&	6245		&	37	\\
205	&	7.3730	&	0.0602	&	0.02500	&	0.00125	&	64	&	16	&	4785		&	4	\\
\midrule
154	&	7.5960	&	0.0493	&	0.02020	&	0.00101	&	64	&	26	&	9162		&	47	\\
167	&	7.5960	&	0.0493	&	0.02020	&	0.00101	&	64	&	24	&	6669		&	37	\\
182	&	7.5960	&	0.0493	&	0.02020	&	0.00101	&	64	&	22	&	7115		&	37	\\
200	&	7.5960	&	0.0493	&	0.02020	&	0.00101	&	64	&	20	&	3017		&	4	\\
222	&	7.5960	&	0.0493	&	0.02020	&	0.00101	&	64	&	18	&	4952		&	8	\\
250	&	7.5960	&	0.0493	&	0.02020	&	0.00101	&	64	&	16	&	7130		&	9	\\
\midrule
153	&	7.8250	&	0.0404	&	0.01640	&	0.00082	&	64	&	32	&	2574		&	8	 \\
163	&	7.8250	&	0.0404	&	0.01640	&	0.00082	&	64	&	30	&	4757		&	24	\\
174	&	7.8250	&	0.0404	&	0.01640	&	0.00082	&	64	&	28	&	14128		&	49	\\
188	&	7.8250	&	0.0404	&	0.01640	&	0.00082	&	64	&	26	&	13911		&	48	\\
204	&	7.8250	&	0.0404	&	0.01640	&	0.00082	&	64	&	24	&	4555		&	7	\\
222	&	7.8250	&	0.0404	&	0.01640	&	0.00082	&	64	&	22	&	5109		&	7	\\
244	&	7.8250	&	0.0404	&	0.01640	&	0.00082	&	64	&	20	&	4433		&	4	\\
271	&	7.8250	&	0.0404	&	0.01640	&	0.00082	&	64	&	18	&	5340		&	4	\\
305	&	7.8250	&	0.0404	&	0.01640	&	0.00082	&	64	&	16	&	6238		&	4	\\
\bottomrule
\end{tabular}
\caption{Input parameters of the lattice ensembles corresponding to light-quark masses $m_l=m_s/20$. We show in every case the temperature in MeV, the bare coupling with its corresponding lattice spacing in fm, the geometry $N_{\sigma}^3\cross N_{\tau}$ and the number of configurations.}
\label{tab:ensembles_ms20}
\end{table}

\begin{table}
\centering
\begin{tabular}{ccccccccc}
\toprule
$T$ [MeV] & $\beta$ & $a$ [fm] &  $a \cross m_s$ & $a \cross m_l$ & $N_{\sigma}$ & $N_{\tau}$ &  \# conf. & streams \\
\midrule
286	&	8.4000	&	0.0247	&	0.008870		&	0.0017740	&	64	&	28	&	4234	&	16	\\
308	&	8.4000	&	0.0247	&	0.008870		&	0.0017740	&	64	&	26	&	4841	&	16	\\
333	&	8.4000	&	0.0247	&	0.008870		&	0.0017740	&	64	&	24	&	4728	&	4	\\
364	&	8.4000	&	0.0247	&	0.008870		&	0.0017740	&	64	&	22	&	5272	&	4	\\
400	&	8.4000	&	0.0247	&	0.008870		&	0.0017740	&	64	&	20	&	5664	&	4	\\
444	&	8.4000	&	0.0247	&	0.008870		&	0.0017740	&	64	&	18	&	6188	&	4	\\
\colorrow
500	&	8.4000	&	0.0247	&	0.008870		&	0.0017740	&	64	&	16	&	5971	&	4	\\
\midrule
\colorrow
195	&	7.5700	&	0.0505	&	0.019730		&	0.0039460	&	64	&	20	&	5911	&	12	\\	
\colorrow
195	&	7.7770	&	0.0421	&	0.016010		&	0.0032020	&	64	&	24	&	5480	&	4	\\	
\colorrow
195	&	8.2490	&	0.0280	&	0.010110		&	0.0020220	&	96	&	36	&	4082	&	4	\\	
\colorrow
220	&	7.7040	&	0.0449	&	0.017230		&	0.0034460	&	64	&	20	&	7933	&	12	\\	
\colorrow
220	&	7.9130	&	0.0374	&	0.014000		&	0.0028000	&	64	&	24	&	5754	&	4	\\	
\colorrow
220	&	8.2490	&	0.0280	&	0.010110		&	0.0020220	&	96	&	32	&	2522	&	2	\\	
\colorrow
251	&	7.8570	&	0.0393	&	0.014790		&	0.0029580	&	64	&	20	&	9443	&	4	\\	
\colorrow
251	&	8.0680	&	0.0327	&	0.012040		&	0.0024080	&	64	&	24	&	5336	&	12	\\	
\colorrow
251	&	8.2490	&	0.0280	&	0.010110		&	0.0020220	&	96  &	28	&	4043	&	4	\\	
\colorrow
293	&	8.0360	&	0.0336	&	0.012410		&	0.0024820	&	64	&	20	&	9287	&	4	\\	
\colorrow
293	&	8.1470	&	0.0306	&	0.011150		&	0.0022300	&	64	&	22	&	9105	&	12	\\	
\colorrow
293	&	8.2490	&	0.0280	&	0.010110		&	0.0020220	&	96	&	24	&	1375	&	3	\\	
\colorrow
352	&	8.2490	&	0.0280	&	0.010110		&	0.0020220	&	96	&	20	&	6167	&	4	\\	
\midrule
352	&	8.1260	&	0.0311	&	0.011380		&	0.0022760	&	64	&	18	&	4214	&	12	\\	
352	&	8.3620	&	0.0255	&	0.009095		&	0.0018190	&	64	&	22	&	3609	&	16	\\	
400	&	8.2763	&	0.0274	&	0.009861		&	0.0019722	&	64	&	18	&	3441	&	16	\\	
400	&	8.6165	&	0.0205	&	0.007174		&	0.0014348	&	64	&	24	&	4097	&	24	\\	
444	&	8.2612	&	0.0278	&	0.010004		&	0.0020008	&	64	&	16	&	4462	&	16	\\	
444	&	8.6376	&	0.0202	&	0.007036		&	0.0014072	&	64	&	22	&	4025	&	16	\\	
500	&	8.5398	&	0.0219	&	0.007703		&	0.0015406	&	64	&	18	&	3617	&	16	\\	
500	&	8.6647	&	0.0197	&	0.006862		&	0.0013724	&	64	&	20	&	4266	&	24	\\	
500	&	8.8815	&	0.0164	&	0.005626		&	0.0011252	&	64	&	24	&	3808	&	24	\\	
1000	&	9.3653	&	0.0110	&	0.003635		&	0.0007270	&	64	&	18	&	1566	&	8
	\\	
1000	&	9.4910	&	0.0099	&	0.003248		&	0.0006496	&	64	&	20	&	2047	&	4
	\\	
1000	&	9.7085	&	0.0082	&	0.002675		&	0.0005350	&	64	&	24	&	1346	&	4
	\\	
10000	&	12.1034	&	0.00110	&	0.0003221	&	0.00006442	&	64	&	18	&	1373	&	8
	\\	
10000	&	12.2281	&	0.00099	&	0.0028855	&	0.00005771	&	64	&	20	&	1479	&	4
	\\	
10000	&	12.4438	&	0.00082	&	0.0023860	&	0.00004772	&	64	&	24	&	943	&	4
	\\	
\bottomrule
\end{tabular}
\caption{Input parameters of the lattice ensembles corresponding to light-quark masses $m_l=m_s/5$. We show in every case the temperature in MeV, the bare coupling with its corresponding lattice spacing in fm, the geometry $N_{\sigma}^3\cross N_{\tau}$ and the number of configurations separated by 10 MC trajectories. The highlighted pink regions correspond to ensembles generated in previous works of the HotQCD collaboration \cite{Altenkort:2023oms,Altenkort:2023eav} or the TUMQCD collaboration~\cite{Bazavov:2019qoo}.
}
\label{tab:ensembles_ms5}
\end{table}

We have performed lattice calculations in (2+1)-flavor QCD with physical strange-quark mass $m_s$ over a temperature range from $T=149$ MeV to $10$ GeV  using the Highly Improved Staggered Quarks (HISQ) action \cite{Follana:2006rc} for the fermion sector and the tree-level improved L\"uscher-Weisz action \cite{Luscher:1984xn,Luscher:1985zq} for the gauge fields. The calculations have been performed on $N_{\sigma}^3\cross N_{\tau}$ boxes for two values of the light-quark mass: $m_l=m_s/20$, which corresponds to the pion mass of $160$ MeV in the continuum limit (and thus is close to the physical point), and $m_l=m_s/5$, corresponding to a pion mass of $320$ MeV. The lattice strange-quark mass was obtained from the parameterization of the line of constant physics (LCP) of \cite{HotQCD:2014kol}. The lattice spacing, and thus the temperature scale, has been fixed through the $r_1$ scale determined in Refs.~\cite{HotQCD:2014kol,Bazavov:2017dsy} with the value $r_1=0.3106(18)$ fm \cite{Bazavov:2010sb}. Though a very recent determination of $r_1$ in physical units yielded a smaller central value \cite{Larsen:2025wvg}, it is still compatible with the scale used in this paper.

The gauge configurations were generated using the usual Rational Hybrid Monte Carlo (RHMC) algorithm \cite{Clark:2003na,Clark:2004cp}, where each configuration is saved after 10 Monte Carlo (MC) trajectories with an acceptance rate tuned to about $80\%$. At each temperature we generated several independent MC streams.

The calculations for the almost physical quark mass ($m_l=m_s/20$) are performed for three values of the bare gauge coupling $\beta$, i.e. three lattice spacings that covers temperatures ranging from $149$ to $305$ MeV, obtaining different temperature by varying $N_{\tau}$. The parameters of these calculations are summarized in Tab.~\ref{tab:ensembles_ms20}. The calculations with $m_l=m_s/5$ span a wide temperature range from $195$ MeV to 10 GeV. This is due to the light-quark masses having a small impact at high temperatures. The parameters of these simulations are summarized in Tab.~\ref{tab:ensembles_ms5}.

We have generated the data sets listed in Tab.~\ref{tab:ensembles_ms20} and~\ref{tab:ensembles_ms5} according to the setup described above. These tables show the bare coupling in each case, the quark masses in terms of lattice spacing, and the number of configurations for each ensemble. Some ensembles (mainly those of $m_l=m_s/5$) were generated directly at some target temperatures, while others were generated over a wider range of temperatures for a given lattice spacing; for the latter, temperature interpolations are required for the continuum extrapolations of Sec.~\ref{subsec:continuum} to ensure the existence of at least three different lattice spacings in all cases. Consequently, we have highlighted in pink the ensembles from previous works of the HotQCD collaboration \cite{Altenkort:2023eav,Altenkort:2023oms} or of the TUMQCD collaboration~\cite{Bazavov:2019qoo}, which are also used in the present analysis. Note that in the present work, as in the previous ones, the bare coupling is defined as $\beta=10/g_0^2$, which is a common notation for improved actions. 

As in Refs.~\cite{Altenkort:2023eav,Altenkort:2023oms}, we use the following discretization of $E_i$ and $B_i$ fields:
\begin{eqnarray}
    E_i(\mathbf{x},\tau) &=& U_i(\mathbf{x},\tau) U_4(\mathbf{x}+\hat{i},\tau) - U_4(\mathbf{x},\tau)U_i(\mathbf{x}+\hat{4})\,,
    \label{eq:e-field}\\
    B_i(\mathbf{x}) &=& \epsilon_{ijk}\left( U_j(\mathbf{x})U_k(\mathbf{x}+\hat{j})-U_k(\mathbf{x})U_j(\mathbf{x}+\hat{k})\right)/2\,.
    \label{eq:b-field}
\end{eqnarray}
As briefly introduced in Sec.~\ref{sec:theoretical-framework}, direct measurements of the $G_{E,B}$ correlators are subject to substantial gauge fluctuations, resulting in a low signal-to-noise ratio. While the multi-level algorithm \cite{Luscher:2001up} successfully mitigates this problem in theories with strictly local interactions, such as quenched QCD, it is not applicable to real QCD with dynamical fermions. In addition, the magnetic correlator requires a renormalization method due to its non-trivial anomalous dimension \cite{Eichten:1990vp}. To address both issues at once, we use the Symanzik-improved GF \cite{Ramos:2015baa}, also known as the "Zeuthen flow", which is a modification of the original proposals \cite{Narayanan:2006rf,Luscher:2009eq,Luscher:2010iy,Luscher:2011bx} by incorporating an additional term that provides an $O(a^2)$ improvement to the flow action. 

Briefly, and just to set the notation, the Zeuthen flow is a 4+1-dimensional local formulation of the gauge sector of the theory, where the original gauge link $U_\mu(\vb{x})$ is replaced by a new set of "flow-time" dependent links $V_\mu(\tau_F,\vb{x})$ subject to the following flow equation:
\begin{eqnarray}
	\left[\partial_{\tau_F}V_\mu(\tau_F,\vb{x})\right]V_\mu^\dagger(\tau_F,\vb{x})&=&-g_0^2\left(1+\frac{1}{12}\nabla^\star_\mu\nabla_\mu\right)\partial_{\vb{x},\mu}S_{\rm{LW}}\left[V\right]
	\label{eq:zeuthen-flow}\,,\\
	V_{\mu}(0,\vb{x})&=&U_\mu(\vb{x})\,,
	\label{eq:zeuthen-intial}
\end{eqnarray}
where $\nabla_\mu$ and $\nabla^\star_\mu$ stand for the usual lattice forward and backward covariant derivatives, respectively, and $S_{\rm{LW}}$ is the usual combination of plaquettes and rectangles of the L\"uscher-Weisz action \cite{Luscher:1984xn,Luscher:1985zq}. Note that the flow time $\tau_F$ has units of length squared. 

The main feature of Eq.~\eqref{eq:zeuthen-flow} is to remove short-range fluctuations by smearing the gauge field over a region of radius $\sqrt{8\tau_F}$, which naturally becomes the inverse of a renormalization scale $\mu_{F}$. In essence, the Zeuthen Flow is, indeed, a renormalization scheme. For our analysis, we will focus on the region where $\sqrt{8\tau_F}/a\geq 1$, making sure that the flow time is always large enough to account for at least one lattice spacing. We also need to keep the flow scale smaller than the relevant physical scales of the calculations, and previous studies have found that an effective criterion is to use flow times such that $\sqrt{8\tau_F}/\tau\leq0.30$ \cite{Altenkort:2020fgs,Brambilla:2022xbd,Eller:2018yje}. This is also consistent with the perturbative NLO calculations of the electric correlator, which predict a linear dependence in $\tau_F$ when the flow time is constrained to $\sqrt{8\tau_F}\leq\tau$ \cite{Eller:2021qpp}. Small values of $\tau T$ are drastically affected by lattice artifacts, and our tests indicate that the amount of flow needed to eliminate these discretization effects is already close to the maximum flow scale $\sqrt{8\tau_F}/\tau = 0.30$. Consequently, we will only consider the range $0.25 \leq \tau T \leq 0.50$ for our extrapolations. Moreover, while we have set the minimum flow time to be at least of one lattice spacing, we have observed in practice that for $\tau T \geq 0.25$ the flow dependence of the correlators is rather small if the flow time is restricted to the range $\sqrt{8\tau_F}/\tau\geq0.25$. Thus, our continuum/flow extrapolations will be performed within this specific range:
\be
	0.25\leq\tau T\leq0.50\text{ and }0.25\leq\sqrt{8\tau_F}/\tau\leq0.30\,,
	\label{eq:extrap_range}
\ee
at a fixed $\tau T$, and in the case of continuum extrapolations, also at a fixed $\sqrt{8\tau_F}/\tau$. 

\begin{figure}[t]
    \centerline{
        \includegraphics[width=0.75\textwidth]{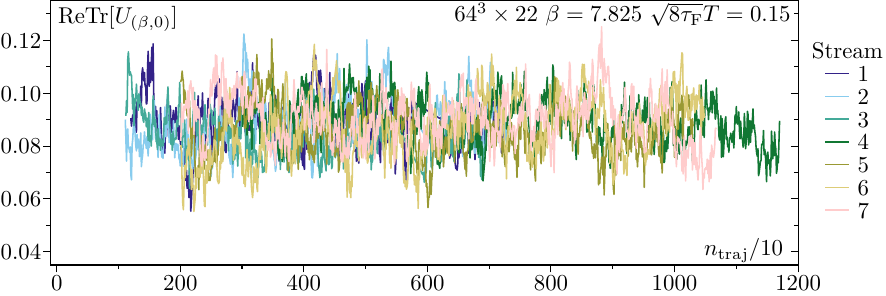}
    }
    \caption{Polyakov loop per stream as a function of the MC trajectories for $\beta=7.8250$ and $N_\tau=22$.}
    \label{fig:polyakov}
\end{figure}

We estimate the autocorrelation time of each stream by analyzing the Polyakov loop evaluated at the maximum value of flow radius used in our extrapolations. For our setup, this occurs at $\sqrt{8\tau_F}/\tau=0.30$ and $\tau T=0.50$, or equivalently, $\sqrt{8\tau_F}T=0.15$. Figure~\ref{fig:polyakov} shows an example of the Polyakov loop as a function of the MC trajectories for one of our ensembles. In most cases, the autocorrelation time is about 100 to 300 MC trajectories. Therefore, we can consider configurations separated by a few hundred MC trajectories to be successfully decorrelated, and use this separation to perform regular binning for bootstrap analysis. In a few rare cases, the data streams are too short to reliably determine the integrated autocorrelation time, and in such cases, we manually truncate and bin the streams to about 300 MC trajectories; this happens only in a handful of streams. 

The lattice generation and GF analysis presented in this work were performed using \textcolor{DeepPink}{SIMULATeQCD} \cite{Mazur:2021zgi,Altenkort:2021cvg,HotQCD:2023ghu} and the \textcolor{DeepPink}{MILC} code (with \textcolor{DeepPink}{SciDAC} libraries) at several high performance computing (HPC) centers around the world. 

\subsection{Normalization and Leading Order Corrections}
\label{subsec:norms}

Since the correlators $G_{E,B}(\tau,T)$ are steep functions with respect to $\tau$, it is useful to normalize them with a theoretically motivated factor in order to understand their $\tau$ dependence, and also to separate lattice artifacts and thermal effects from the vacuum contribution. We will use different normalization factors for different purposes: some to illustrate the behavior of the correlators, and others to mitigate lattice effects.

The first normalization approach is obtained by using the leading order (LO) pQCD results for the spectral functions. We will leave the technical details to be discussed in Sec.~\ref{sec:spectral}, but for both, the electric and magnetic cases, the
LO result is \cite{Caron-Huot:2009ncn,Bouttefeux:2020ycy}:
\be
	\rho_{E,B}^{\LO}(\omega,\tau)=\frac{g^2_{\MS}(\mu) C_F}{6 \pi}\omega^3\,,
	\label{eq:rho_LO}
\ee
where $C_F = (N_c^2-1)/2N_c$ is the Casimir factor, and $g^2_{\MS}(\mu)$ is the running coupling in the $\MS$ scheme at scale $\mu$. The transport contributions, proportional to  $\kappa_E$ or $\kappa_B$, do no contribute at this order as $\kappa_{E,B}$ starts at order $g^4$.
Then, the LO correlation function obtained by integrating Eq.~\eqref{eq:current-current-correlation} is \cite{Caron-Huot:2009ncn}: 
\be
    \nonumber
	G^{\LO}(\tau,T) = g^2_{\MS}(\mu)C_F G^{\rm{norm}}(\tau,T)\,,
\ee
with
\be
    G^{\rm{norm}}(\tau,T) \equiv \pi^2T^4\left[\frac{\cos^2(\pi\tau T)}{\sin^4(\pi\tau T)}+\frac{1}{3\sin^2(\pi\tau T)}\right]\,.
	\label{eq:GLO_norm}
\ee
By dividing our lattice results by $G_{\rm norm}$, we take care of the leading order $\tau$ dependence of the correlator (up to an overall constant $g^2 C_F$). The $\tau$ dependence due to higher order perturbative corrections, such as running of the coupling constant and the non-zero value of $\kappa_{E,B}$, will contribute to the $\tau$ dependence of the correlator as will be discussed below. 

In Refs.~\cite{tuprints23185,Altenkort:2023oms,Altenkort:2024spl}, the LO results for the chromo-electric and chromo-magnetic correlators have been calculated on the lattice, at finite flow times, for the improved gauge action used in this study, and for the simple discretizations of the fields given by Eqs.~\eqref{eq:e-field} and \eqref{eq:b-field}. As in our previous work \cite{Altenkort:2023oms,Altenkort:2024spl}, we will use tree-level (LO) improvement to reduce the lattice discretization effects, as well as the distortions due to GF, by replacing the original lattice data $G_{E,B}^{\rm{data}}(\tau,T)$ by
\be
	G_{E,B}(\tau,T)\equiv G_{E,B}^{\rm{data}}(\tau,T)\cross\frac{G^{\rm{norm}}(\tau,T)}{G_{E,B}^{\rm{latt}}(\tau, T,N_{\tau},\tau_F) }\,,
	\label{eq:final_norm_G}
\ee
where $G_{E,B}^{\rm{latt}}(\tau T,N_{\tau},\tau_F)$ is the analog of $G^{\rm{norm}}(\tau,T)$ calculated at non-zero lattice spacings and finite flow time for the discretizations used in this study. As we shall see, the lattice artifacts and flow distortions present in $G_{E,B}^{\rm{data}}$ are largely removed by dividing by $G_{E,B}^{\rm{latt}}(\tau, T,N_{\tau},\tau_F)$. The remaining artifacts are removed by continuum and flow time extrapolations.

\subsection{Interpolations and continuum extrapolations}
\label{subsec:continuum}

\begin{figure}[t]
    \centerline{
        \includegraphics[width=0.5\textwidth]{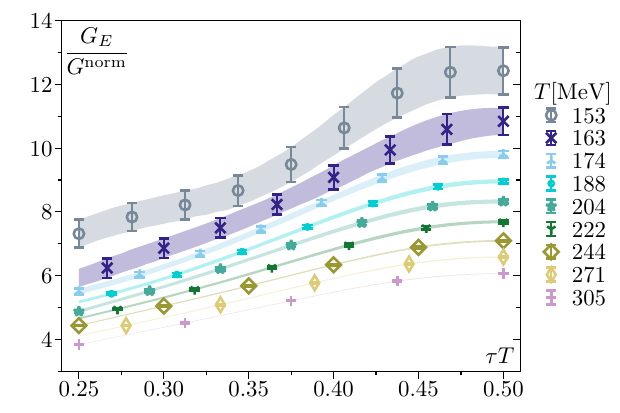}
        \includegraphics[width=0.5\textwidth]{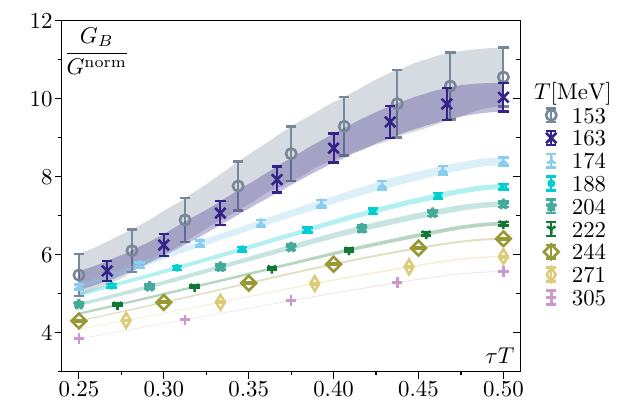}
    }
    \caption{The chromo-electric (left) and chromo-magnetic (right) correlators as function of $\tau T$ at  $\beta=7.8250$ and $\sqrt{8 \tau_F}/\tau=0.3$ for different temperatures. }
    \label{fig:interpolation_tau}
\end{figure}

In Fig.~\ref{fig:interpolation_tau}, we show as an example our lattice results for chromo-electric and chromo-magnetic correlators, divided by $G_{\rm norm}$, at $\beta=7.825$ and for a relative flow time $\sqrt{8 \tau_F}/\tau=0.3$. As one can see from the figures, the data follow the pattern seen in the previous calculations, namely $G_{E,B}/G_{\rm norm}$ decreases with increasing temperature and shows mild $\tau$ dependence. At $T=153$ MeV, the correlators are very noisy and therefore will not be used to obtain the heavy-quark diffusion coefficient.

The first step in extracting the diffusion coefficient is to perform a continuum extrapolation of $G_{E,B}/G_{\rm norm}$ at a fixed relative flow time. However, since we have different $\tau$ values at different lattice spacings, we must first interpolate in $\tau$, so that we can extrapolate on a given set of temporal site values. We use cubic splines for these interpolations, and an example for $\beta=7.825$ can be found in Fig.~\ref{fig:interpolation_tau}. 

\begin{figure}[t]
    \centerline{
        \includegraphics[width=0.5\textwidth,page=21]{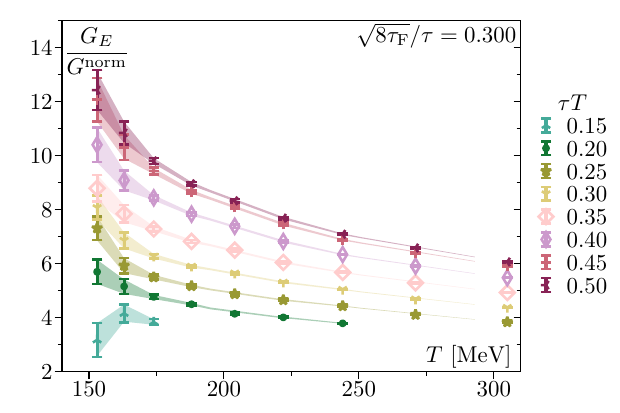}
        \includegraphics[width=0.5\textwidth,page=21]{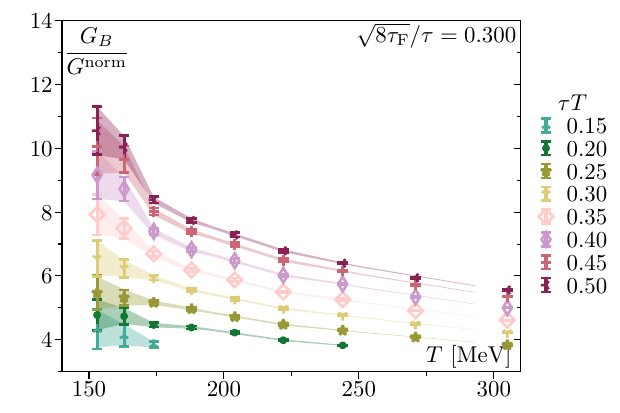}
    }
    \caption{Temperature interpolation of $m_l=m_s/20$ electric and magnetic correlator data at fixed relative flow time $\sqrt{8\tau_F}/\tau=0.30$ for $\beta=7.8250$. The color bands are linear splines and the points correspond to the data obtained for parameters shown in Tab.~\ref{tab:ensembles_ms20}. }
    \label{fig:interpolation_EE_BB}
\end{figure}

Unlike in the previous calculations, where we had three lattice spacings at each temperature value, in this study we typically vary the temperature by varying $N_{\tau}$ at a fixed lattice spacing. Only in the high-temperature region we have three lattice spacings for a given temperature value. Therefore, we also need to perform interpolations in the temperature in most cases, and we find that it is relatively easy to perform them in terms of $G_{E,B}/G_{\rm norm}$. In Fig.~\ref{fig:interpolation_EE_BB}, we show linear interpolations on the temperature for $G_{E,B}/G_{\rm norm}$ at $\beta=7.825$, as this data set is the most challenging. 

Since in this study we use two different light-quark masses, the quark-mass dependence of the correlators should also be addressed. As mentioned above, the light-quark mass effects are expected to be different in the low- or high-temperature regions, and for these reasons, in what follows we discuss the lattice-spacing and quark-mass dependence separately for three temperature regions: the low-temperature region, $T<205$ MeV, the intermediate-temperature region, $220 ~{\rm MeV}\le T \le 308 ~\rm MeV $, and the high-temperature region, $T>352$ MeV.

\subsubsection{Interpolations and continuum extrapolations of correlation functions in the low-temperature region}
\label{subsubsec:continuum_low}

For $T<205$ MeV, we have results for $m_l=m_s/20$ at three lattice spacings corresponding to  $\beta=7.373,~7.596$ and $7.825$. Since the data for $T=153$ MeV and $\beta=7.825$ are too noisy, the lowest temperature for which we perform continuum extrapolation is $T=163$ MeV. For $\beta=7.373$ and $7.596$, we use all the temperatures shown in Tab.~\ref{tab:ensembles_ms20} to interpolate to $T=163$ MeV. Additionally, we perform temperature interpolations to the following values: $T=174,~188$ and $195$ MeV. 

\begin{figure}
    \centerline{
        \includegraphics[width=0.5\textwidth,page=11]{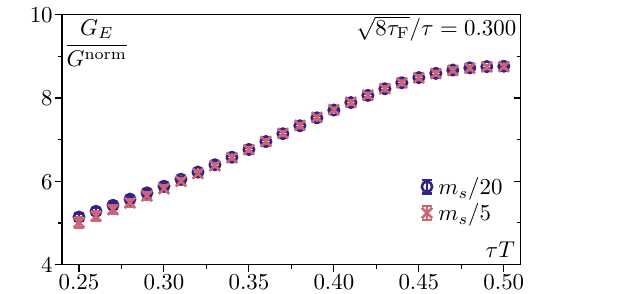}
        \includegraphics[width=0.5\textwidth,page=11]{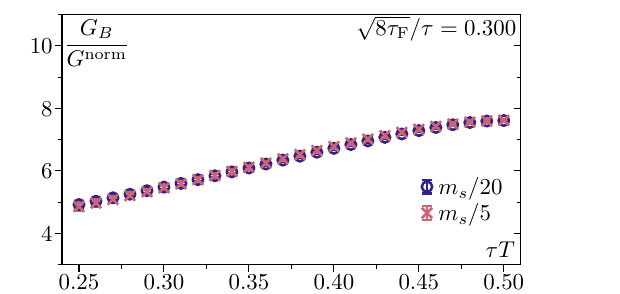}
    }
    \caption{The $\tau$-dependence of $G_E/G_{\rm norm}$ (left) and $G_B/G_{\rm norm}$ (right) at $T=195$ MeV for $\sqrt{8 \tau_F}/\tau=0.3$. }
    \label{fig:comp_T195}
\end{figure}

Then, we perform continuum extrapolations at these temperatures. Based on previous perturbative studies \cite{Altenkort:2024spl,Altenkort:2023eav,Altenkort:2023oms,Brambilla:2019oaa,Brambilla:2020siz}, we expect that the slopes of the extrapolations will approach zero for very large $\tau T$, with a leading term proportional to $1/(\tau T)^2$. Therefore, we will perform combined extrapolations for all $\tau T$, at a fixed relative flow time, using the following Ansatz:
\be
	G_{E,B}(\tau T,a/r_1)=G_{E,B}^{\rm{cont}}(\tau T)-\left(\frac{b}{\tau T}\right)^2\cross\left(\frac{a}{r_1}\right)^2\,,
	\label{eq:continuum_ansatz}
\ee
where $G_{E,B}^{\rm{cont}}(\tau T)$ is the continuum value and $(b/\tau T)^2$ is the slope of the extrapolation. This combined fit improves the statistical errors by simultaneously fitting all data for a given $\tau T$ and $\sqrt{8\tau_F}/\tau$. Some examples of continuum extrapolations are shown in the Appendix~\ref{appendix:extrapolation}. For $T=195$ MeV, we calculated in previous works $G_E$ and $G_B$ for $m_l=m_s/5$. Therefore, it make sense to compare the calculations at these two light-quark masses schemes. This comparison is shown in Fig.~\ref{fig:comp_T195} for $\sqrt{8 \tau_F}/\tau=0.3$. As one can see from the figure, there are no statistically significant effects of the light-quark mass at $T=195$ MeV. 

\subsubsection{Interpolations and continuum extrapolations of correlation functions in the intermediate-temperature region}
\label{subsubsec:continuum_inter}

In the intermediate-temperature region, we have previously performed calculations at $T=220,~251$ and $293$ MeV with $m_l=m_s/5$. We have results in this region with $m_l=m_s/20$ with either one or two lattice spacings, as well as lattice results with $m_l=m_s/5$ for $\beta=8.4$ ($a=0.025$ fm) at two temperatures. As we have seen above, at $T=195$ MeV, the $m_l/20$ and $m_s/5$ results for the chromo-electric and chromo-magnetic correlators agree within errors. Thus, we may expect the same in the intermediate-temperature region, so it makes sense to combine both lattice results, $m_l=m_s/5$ and $m_l=m_s/20$, for the continuum extrapolations, which increases the number of lattice spacings available in each case. We find that combined continuum extrapolations work well. In addition, we can use the lattice results for $G_E$ and $G_B$ for two temperatures at $\beta=8.4$ to interpolate to $T=293$ MeV. With all these ensembles, we have five different lattice spacings at our disposal for continuum extrapolations in this temperature region. Consequently, we perform those extrapolations using Eq. (\ref{eq:continuum_ansatz}), and some examples can be found in Appendix~\ref{appendix:extrapolation}.

\subsubsection{Interpolations and continuum extrapolations of correlation functions in the high-temperature region}
\label{subsubsec:continuum_high}

The highest temperature in the previous HotQCD lattice studies of the diffusion coefficient was $T=352$ MeV \cite{Altenkort:2023oms,Altenkort:2024spl}. However, the calculations at this temperature have been performed at a single value of the lattice spacing, $a=0.028$ fm. Therefore, in the present work, we performed calculations of $G_E$ and $G_B$ at two additional lattice spacings for $T=352$ MeV, see Tab.~\ref{tab:ensembles_ms5}. Additionally, using the results for $\beta=8.4$, we can perform interpolations to obtain $G_E$ and $G_B$ at $T=352$ MeV with the smallest lattice spacing used in this study, $a=0.025$ fm. Therefore, altogether we have four lattice spacings for the continuum extrapolations at this temperature. Furthermore, in the present study, we extended the temperature range to higher values, considering $T=400$, $444$, $500, 1000$ and $10000$ MeV. At each one of these temperatures, we performed lattice QCD calculations at three lattice spacings. Then, we used Eq.~\eqref{eq:continuum_ansatz} again for the continuum extrapolations in this temperature region, and some examples can be found in Appendix~\ref{appendix:extrapolation}.

\subsection{Renormalization and flow extrapolation}
\label{subsec:flow}

The chromo-electric and chromo-magnetic correlators have to be renormalized before the spectral function can be extracted from them. The renormalization constants of these correlators, $Z_E$ and $Z_B$, are scheme and scale dependent. On the lattice, these renormalization constants in general have to be calculated before taking the continuum limit \cite{Christensen:2016wdo,Banerjee:2022uge}. Using the GF when calculating $G_E$ and $G_B$, however, automatically gives the renormalized correlation functions in the GF scheme at scale $\sqrt{8 \tau_F}$, up to corrections proportional to powers of $\sqrt{8 \tau_F}/\tau$. To remove these corrections (artifacts), we need to perform the $\tau_F\rightarrow 0$ extrapolation.
For the chromo-electric correlator this extrapolation is straightforward, since at 1-loop this correlator is the same in the GF and $\overline{\mathrm{MS}}$ schemes and is scale independent in each of these schemes.
This is not the case for the chromo-magnetic correlator, and the strategy of the $\tau_F\rightarrow 0$ extrapolation is discussed below.

\subsubsection{\texorpdfstring{$G_B$}{} matching factors}
\label{subsubsec:matching}

\begin{figure}[t]
    \centerline{
        \includegraphics[width=0.5\textwidth]{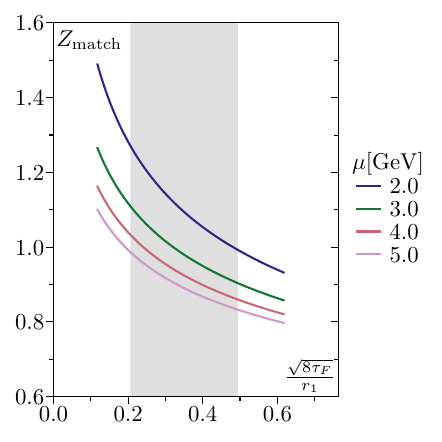}
    }
    \caption{Matching factor $Z_{\rm{match}}$ as a function of $\sqrt{8\tau_F}/r_1$, for several choices of the $\MS$ scale $\mu$. The gray band in the plot represents the range of flow times considered for the extrapolation $\tau_F\rightarrow0$, as specified in Eq.~\eqref{eq:extrap_range}.}
    \label{fig:matching_factor}
\end{figure}

The chromo-magnetic correlation function is scheme and scale dependent already at 1-loop level. Since all the perturbative calculations of the chromo-magnetic spectral function used in this work are in the $\MS$ scheme \cite{Bouttefeux:2020ycy}, it is important to know the matching between the GF and $\MS$. This matching has been calculated at 1-loop \cite{delaCruz:2024cix,Brambilla:2023vwm}:
\be
	G_B^{\MS}(\tau T,\mu)=G_B^{\rm{GF}}(\tau T,\mu_{F})\left[1-\gamma_0g_{\MS}^2(\mu)\left(\ln\frac{\mu^2}{4\mu_F^2}+\gamma_{\rm{E}}\right)\right]\,,
	\label{eq:GFtoMS_leading}
\ee
where $\gamma_0=3/(8\pi^2)$ is the leading anomalous dimension and $\gamma_{\rm{E}}=0.5772156649$ is the Euler-Mascheroni constant. To improve the conversion and to account for unknown higher-loop effects and multiple logarithmic corrections, we can extend Eq.~\eqref{eq:GFtoMS_leading} to \cite{Altenkort:2024spl}:
\begin{eqnarray}
	G_B^{\MS}(\tau T,\mu)&=&Z_{\rm{match}}(\mu,\mu_{F})G_B^{\rm{GF}}(\tau T,\mu_{F})\,,
	\label{eq:GFtoMS}\\
	\ln Z_{\rm{match}}(\mu,\mu_{F})&=&-\gamma_0g_{\MS}^2(\mu)\left(\ln\frac{\mu^2}{4\mu_F^2}+\gamma_{\rm{E}}\right)\,.
\end{eqnarray}

In this study, we convert the continuum extrapolated results for $G_B^{\rm{GF}}(\tau T,\mu_{F})$ to the $\MS$ scheme using the above equations, and then perform the $\tau_F$ extrapolation. Unlike Ref.~\cite{Altenkort:2024spl}, we do not use the physical scheme for $G_B$ introduced in \cite{Laine:2021uzs,Bouttefeux:2020ycy}, since comparison with weak coupling calculations and the extraction of $\kappa_B$ can be entirely performed within the $\MS$ scheme. However, an extra step is needed to obtain the final physical value $\kappa_B$ from these correlators. We will discuss this in more detail in Sec.~\ref{subsec:spectral_B}.

For the numerical calculation of the matching factor, we use the $g_{\MS}^2$ perturbative results up to five loops calculated with RunDec package, with $\LMS=0.339$ GeV and $N_f=3$. This way to evaluate $g_{\MS}^2$ will be used throughout this paper. We need to choose the $\MS$ renormalization scale $\mu$ such that the logarithmic corrections do not become too large. In this study, and for temperatures $T\neq 10$ Gev, we consider four choices: $\mu=2$, $3$, $4$, and $5$ GeV, and find that for these values the logarithmic corrections are never too large to be of concern. For the highest temperature $T=10$ GeV, the logarithmic corrections are already large, so only for this temperature, we will consider a higher $\MS$ scale of $\mu=50$ GeV. Fig.~\ref{fig:matching_factor} shows the matching factor $Z_{\rm{match}}(\mu,\mu_{F})$ for these four choices of $\mu$, plotted as a function of $\sqrt{8\tau_F}/r_1$. The change in $\mu$ mostly amounts to the overall shift in the value of the matching coefficient, and does not change its $\tau_F$-dependence in the flow time window specified in Eq.~\eqref{eq:extrap_range}, which is indicated by the gray band in Fig.~\ref{fig:matching_factor}. As a consequence, the curvature of $G_B^{\rm{GF}}(\tau T,\mu_{F})$, which encodes the value of $\kappa_B$, remains practically unchanged by a global rescaling such that $\kappa_B$ turns out as $\mu$-independent within errors.

\subsubsection{Flow extrapolation}
\label{subsubsec:flow_extrap}

\begin{figure}[t]
    \centerline{
        \includegraphics[width=0.4\textwidth]{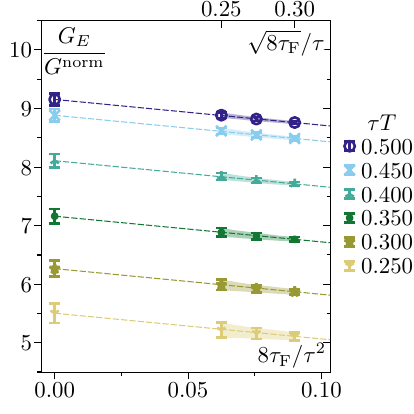}
        \includegraphics[width=0.4\textwidth]{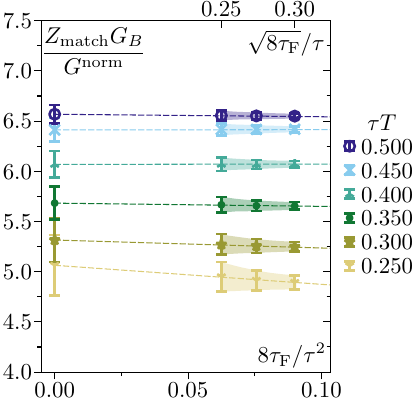}
    }
    \centerline{
        \includegraphics[width=0.4\textwidth]{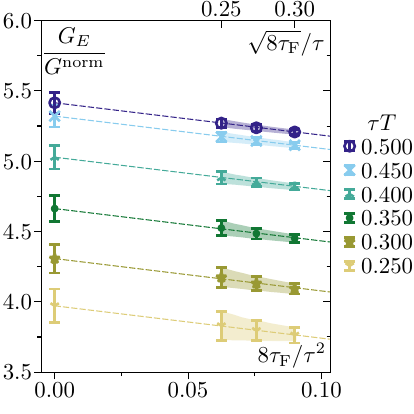}
        \includegraphics[width=0.4\textwidth]{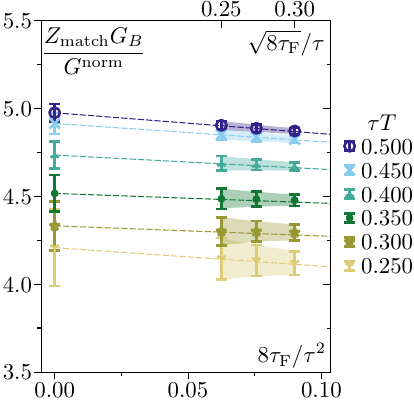}
    }
    \caption{Flow extrapolation of electric and magnetic correlators. Upper left: electric correlator for $T=195$ MeV. Upper right: magnetic correlator for $T=195$ MeV at $\mu=4$ GeV. Lower left: electric correlator for $T=400$ MeV. Lower right: magnetic correlator for $T=400$ MeV at $\mu=4$ GeV. The dashed lines represent the median values from the fits performed following Eq.~\eqref{eq:flow_ansatz}.}
    \label{fig:flow_extrap}
\end{figure}

\begin{figure}[t]
    \centerline{
        \includegraphics[width=0.5\textwidth]{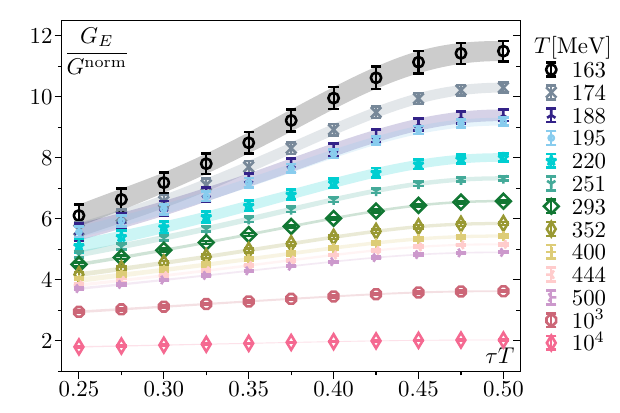}
        \includegraphics[width=0.5\textwidth]{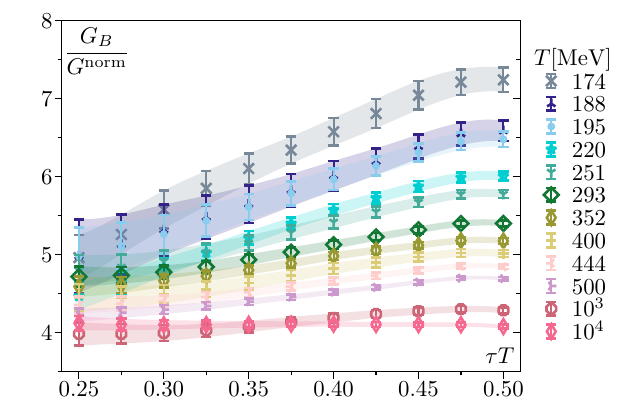}
    }
    \caption{Continuum and flow extrapolated correlators, normalized by Eq.~\eqref{eq:final_norm_G}, as a function of $\tau T$ for different temperatures. The magnetic correlator is evaluated at $\mu=4$ GeV for $T\neq10$ GeV, and at $\mu=50$ GeV for $T=10$ GeV.}
    \label{fig:con_flow_correlator}
\end{figure}

Once the appropriate matching factors have been calculated, we can safely perform the flow extrapolations $\tau_F\rightarrow0$.  Perturbative calculation suggests that the leading flow time dependence of the chromo-electric and chromo-magnetic correlators is linear in $\tau_F$ \cite{Eller:2018yje,Eller:2021qpp,delaCruz:2024cix}. Therefore, we will use linear form to perform the extrapolations. As discussed in Sec.~\ref{sec:numerical-results}, we will restrict the analysis within the range specified in Eq.~\eqref{eq:extrap_range}. Within this range, the signal-to-noise ratio remains robust while the lattice artifacts, which are typically pronounced at small flow times, are effectively constrained. Additionally, it ensures a reasonable linear behavior of the extrapolation across all data sets.

For the electric correlator, the slopes of the extrapolations are expected to have a very weak dependence on $\tau T$ \cite{Eller:2021qpp}, so we choose a combined fit for all $\tau T$ at the same time. On the other hand, for the magnetic correlator, the data clearly shows that the slopes depend on $\tau T$, so we rather use independent extrapolations. In both cases we use the following approach:
\be
	G_{E,B}(\tau T,\tau_F)=G_{E,B}^{\rm{\tau_F\rightarrow0}}(\tau T)+b_{E,B}\cross\tau_F\,,
	\label{eq:flow_ansatz}
\ee
where the slopes $b_{E,B}$ are (independent) dependent of $\tau T$ for the (electric) magnetic correlator.

Fig.~\ref{fig:flow_extrap} shows the flow extrapolation of $G_{E,B}$ for two different temperature values: $T=195$ and $400$ MeV. For the magnetic correlator, we show the extrapolation at the $\MS$ scale $\mu=4$ GeV. For most temperatures, the choice of scale acts only as a multiplicative factor, and the slopes of the extrapolation remain quite independent of $\mu$. As discussed in Sec.~\ref{subsec:norms}, this will lead to a very weak dependence of the diffusion coefficient on the choice of the renormalization scale, since $\kappa_B$ is mainly determined by the curvature of the correlators. In both, the electric and the magnetic cases, the extrapolations give reasonable $\chi^2/\text{d.o.f}$ over the whole temperature range we consider. 

Finally, the continuum and zero flow extrapolated results for both correlators are shown in Fig.~\ref{fig:con_flow_correlator}, normalized using Eq.~\eqref{eq:final_norm_G}, for the full range of temperatures considered in this work. The magnetic correlator is shown at the $\MS$ scale $\mu=4$ GeV. The continuum and flow time extrapolated results of $G_E/G_{\rm norm}$ and $G_B/G_{\rm norm}$ have the same features as the raw lattice results, namely these ratios decrease with increasing temperature and their dependence on $\tau T$ also decreases. At the highest temperature considered, these ratios are almost $\tau$-independent. While continuum or flow extrapolations do not change the general properties of the correlators, the statistical errors increase significantly in the zero flow limit. In the right panel, we do not show  the magnetic correlators for $T=163$ MeV due to the large errors, although we will still use it for the spectral reconstruction of Sec.~\ref{sec:spectral}.

\section{Perturbative results on the chromo-electric and chromo-magnetic correlators}
\label{sec:perturbative_results}

In this section, we review the perturbative results on the chromo-electric and chromo-magnetic correlators, and discuss their comparison with the lattice data. This analysis serves as an important step for the determination of the heavy-quark diffusion coefficient of Sec.~\ref{sec:spectral}.

\subsection{Chromo-electric correlators and spectral functions at NLO}
\label{subsec:kappa_E_PT}

\begin{figure}[t]
    \centerline{
        \includegraphics[width=0.5\textwidth]{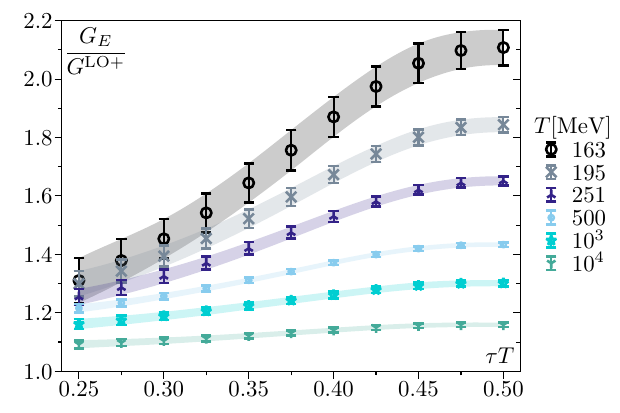}
        \includegraphics[width=0.5\textwidth]{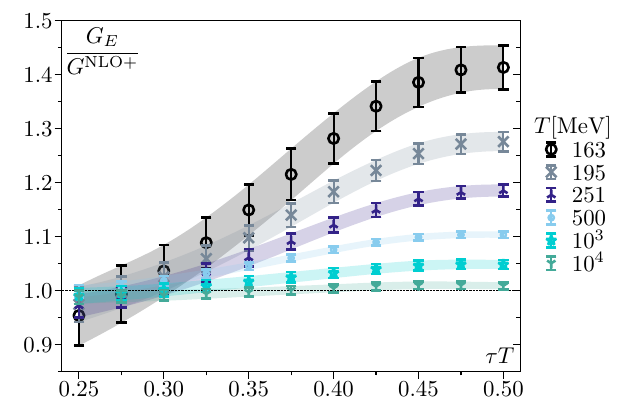}
    }
    \caption{Continuum and flow extrapolated electric correlators, normalized by $G^{\LOp}$ and $G^{\NLOp}$, as a function of $\tau T$ for different temperatures.}
    \label{fig:con_flow_correlator_LOp}
\end{figure}

The chromo-electric correlator has been calculated in the literature at next-to-leading order (NLO) \cite{Burnier:2010rp,delaCruz:2024cix,Scheihing-Hitschfeld:2023tuz}. The full NLO result has no simple analytic form in general. However, it has a simple form for $\omega \ll T$ and for $\omega\gg T$ \cite{Burnier:2010rp}. For $\omega \ll T$ we have \cite{Burnier:2010rp}:
\be
	\rho^{\NLO}_{E}(\omega,T) \simeq \rho^{\rm{IR}}_{E}(\omega,T)=\frac{\kappa_{E}(T)}{2T}\omega\,,
	\label{eq:ir_rho}
\ee
with
\be
    \kappa_E=	\kappa_E^{\rm LO} = \frac{g^4 C_F T^3}{18\pi}\left(\left[N_c+\frac{N_f}{2}\right]\left[\ln\frac{2T}{m_D}+\xi\right]+\frac{N_f\ln2}{2}\right)\,,
    \label{eq:kappa_LO}
\ee
where $\xi=1/2-\gamma_E+\zeta^\prime(2)/\zeta(2)\simeq-0.64718$ and $m_D^2\equiv g^2T^2(N_c+N_f/2)/3$ is the LO Debye mass. The leading order result for $\kappa_E$ was first obtained in Ref.~\cite{Moore:2004tg}. Thus, the NLO result for $\rho_E$ yields the LO result for $\kappa_E$.
For $\omega \gg T$, we have:
\begin{eqnarray}
	\rho_{E}^{\NLO}(\omega,T)\simeq \rho_E^{\rm{UV},\NLO}(\omega)&=&\frac{g^2_{\MS}(\mu) C_F}{6 \pi}\omega^3\left\{1+\frac{g^2_{\MS}(\mu)}{(4\pi)^2}\left[-N_f\left(\frac{2}{3}\ln\frac{\mu^2}{4\omega^2}+\frac{20}{9}\right)\right.\right.
	\label{eq:rho_NLO_E}\\
	\nonumber
	&&\left.\left.{} + N_c\left(\frac{11}{3}\ln\frac{\mu^2}{4\omega^2}+\frac{149}{9}-\frac{2\pi^2}{3}\right)\right]\right\}\,,
\end{eqnarray}
where $g^2_{\MS}(\mu)$ is the running coupling in the $\MS$ scheme at scale $\mu$, and $N_c=3$, $N_f=3$ are the number of colors and quark flavors, respectively. In this limit, the spectral function is independent of the temperature. There was a mistake in the original calculation of $\rho_E$ in Ref.~\cite{Burnier:2010rp}, which was corrected in Refs.~\cite{delaCruz:2024cix,Sambataro:2023tlv}. In the original calculation, the last term in the above equation had been reported as $-\frac{8\pi^2}{3}$. We also note that $\rho_E^{\NLO}$ is independent of $\mu$ up to higher order terms, as the explicit dependence on $\mu$ through the logarithmic correction is canceled by the $\mu$-dependence of $g^2_{\MS}(\mu)$. 

A natural choice of $\mu$ is the one that minimizes the NLO correction in Eq.~\eqref{eq:rho_NLO_E}. Thus, based on the literature, we will refer to such scale as $\mu_{\rm{opt}}$, and is given by:
\be
	\mu_{\rm{opt}}=2\omega\exp\left[\frac{(6\pi^2-149)N_c+20N_f}{6(11N_c-2N_f)}\right]\,.
	\label{eq:omega_uv_opt}
\ee
For $N_c=3$, $N_f=3$, we obtain $\mu^{\NLO}(\omega)\simeq0.549\omega$. Note that, due to the above-mentioned error in the NLO spectral function, this scale changes drastically from its previous value of $\mu^{\NLO}(\omega)\simeq14.7427\omega$ used in Refs.~\cite{Altenkort:2023oms,Altenkort:2024spl}.

The NLO spectral function interpolates smoothly between the limiting cases given by Eq.~\eqref{eq:ir_rho} and Eq.~\eqref{eq:rho_NLO_E} \cite{Burnier:2010rp}, at least for sufficiently high temperature, where the NLO result is trustworthy, see e.g. discussion in Ref.~\cite{Brambilla:2020siz}. For $\omega \sim T$, the choice $\mu=\mu^{\rm{opt}}$ may not be suitable, as the spectral function will depend on $T$. It was argued that a reasonable choice of scale in this case could be the one appearing in the effective theory of QCD at high temperatures, called EQCD \cite{Kajantie:1997tt}:
\be
	\mu_{T}=4\pi T\exp\left[-\gamma_E - \frac{N_c-8N_f\ln2}{22N_c-4N_f}\right]\,.
	\label{eq:mu_T_e}
\ee
For $N_c=3$, $N_f=3$, we have $\mu_{T}\simeq 9.082T$. 

While at NLO the spectral function does not depend on $\mu$, higher order corrections related to the running of the coupling are numerically important, as pointed out in Ref.~\cite{Brambilla:2020siz}. Therefore, it makes sense to use the NLO result for the spectral function with the running coupling to construct the reference correlator for $G_E$, instead of $G_{\rm norm}$ introduced in Sec.~\ref{subsec:norms}. Therefore, we define:
\be
    G_E^{\NLOp}=\int_0^{\infty} \mathrm{d}\omega\ \rho_E^{\rm{UV},\NLO}(\omega) K(\omega,\tau)\,,
\ee
where $\rho_E^{\rm{UV},\NLO}$ is evaluated using Eq.~\eqref{eq:rho_NLO_E} with 
\be
	\mu=\mu^{\NLO}(\omega)\equiv\max(\mu_{\rm{opt}},\mu_{T})\,. 
	\label{eq:omega_uv_thermal_NLO}
\ee
With this normalization, the effect of the running coupling is taken into account. For the numerical evaluation of this quantity, we use the 5-loop running of $g^2_{\MS}(\mu)$. Similarly, we define $G^{\LOp}$ by replacing $\rho_E^{\rm{UV},\NLO}$ with $\rho_E^{\LO}$. Due to the lack of an "optimal" scale in the LO case, we simply choose:
\be
	\mu=\mu^{\LO}(\omega)\equiv\max(\omega,\mu_{T})\,. 
	\label{eq:omega_uv_thermal_LO}
\ee

In Fig.\ref{fig:con_flow_correlator_LOp}, we show the continuum and flow-time extrapolated chromo-electric correlators normalized by these LO and NLO perturbative results, $G^{\LOp}$ and $G^{\NLOp}$, respectively. For clarity, only a subset of temperatures is displayed in the figure. We note that these normalized ratios are smaller than the corresponding $G_E/G^{\rm norm}$ ratios shown in Fig.~\ref{fig:con_flow_correlator}, and they exhibit a different $\tau$ dependence. On one side, the temperature dependence of the ratios at large $\tau$ is substantial, indicating that it is sensitive to the diffusion coefficient $\kappa_E$. In contrast, at small $\tau$, the temperature dependence is much weaker, and furthermore the ratio $G_E/G_E^{\NLOp}$ approaches one, while $G_E/G_E^{\LOp}$ remains slightly above one. This behavior demonstrates that the perturbative calculation accurately captures the behavior of $G_E$ at small $\tau$, as expected, and that the inclusion of NLO corrections improves upon the LO result in this region. Previous studies did not observe such an agreement with perturbative predictions \cite{Brambilla:2020siz,Altenkort:2023oms} due to the mistake in the original NLO expression for $\rho_E$ given in Ref.~\cite{Burnier:2010rp}.

\subsection{Higher order corrections and lattice-to-perturbative comparisons for the electric correlator at high temperature}
\label{subsec:kappa_E_high_T}

\begin{figure}[t]
    \centerline{
        \includegraphics[width=0.5\textwidth]{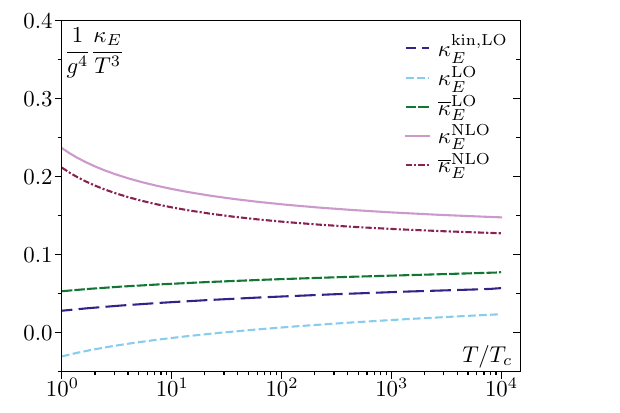}
    	\includegraphics[width=0.5\textwidth]{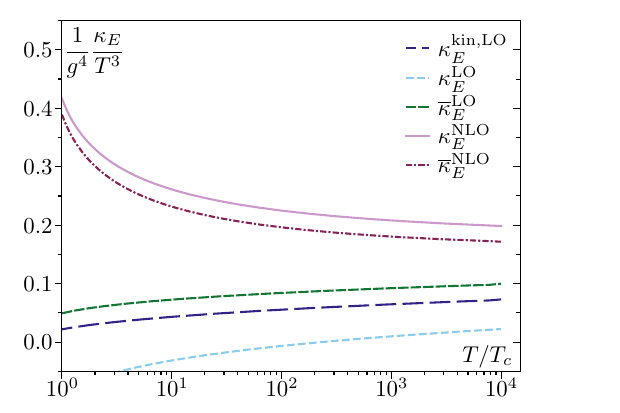}
    }
    \caption{Perturbative estimates of the diffusion coefficient obtained with several mathematical methods. For better visibility, the y-axis has been rescaled by a factor of $1/g^4$. We show the total LO and NLO calculations, along with the resummed results obtained by expanding the gluon self-energy. Left: quenched QCD. Right: (2+1)-flavor QCD.}
    \label{fig:delta_kappa}
\end{figure}

In the previous sub-section, we performed the comparison of the lattice results on the chromo-electric correlator with the perturbative results, obtained from the high $\omega$ part of $\rho_E(\omega,T)$. The comparison with the complete NLO result for $G_E$, however, is problematic unless the temperature is asymptotically high. This is because the LO expression for $\kappa_E$ in Eq.~\eqref{eq:kappa_LO} is obtained under the assumption $m_D \ll T$, and it is negative even at temperatures much larger than the chiral crossover temperature.

\begin{figure}[t]
    \centerline{
        \includegraphics[width=0.5\textwidth]{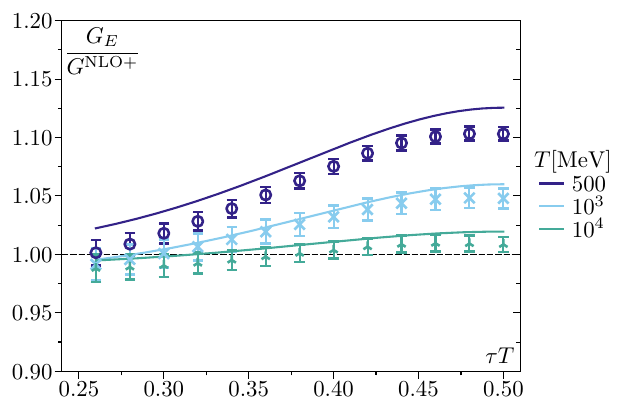}
    }
    \caption{Comparison between the  perturbative-predicted (lines) and the lattice-computed correlators (points) at high temperatures.}
    \label{fig:NLO_kappa}
\end{figure}

At leading order, one can use the kinetic theory to obtain the following expression for $\kappa_E$ \cite{Moore:2004tg}:
\begin{align}
	\label{eq:kappa_kinetic_integral}
	\kappa_E^{\rm kin,LO} \equiv {} & 
    \frac{g^4 C_F}{12\pi^3}\int_0^\infty \mathrm{d}q\,q^2\int_0^{2q}\mathrm{d}p\frac{p^3}{(p^2+\Pi_{00}(p,T))^2} \times \\ \nonumber
	& \left[
		N_f\,n_F(q)[1-n_F(q)]\,\left(2-\frac{p^2}{2q^2}\right)
		+ N_c\,n_B(q)[n_B(q)]\,\left(2-\frac{p^2}{q^2}+\frac{p^4}{4q^4}\right) \right]\,,
\end{align}
where $n_F(q)$ and $n_B(q)$ are the Fermi-Dirac and Bose-Einstein distributions, respectively, and $\Pi_{00}(p,T)$ is the temporal component of the gluon self-energy. For small momenta, $\Pi_{00(p,T)}^\LO\simeq m_D^2$ and, assuming $m_D\ll T$, the integrals in the above equation can be done analytically to reproduce Eq.~\eqref{eq:kappa_LO}. This is shown with label $\kappa_E^{\rm LO}$ in Fig.~\ref{fig:delta_kappa}. However, if we perform the integrals in the above equation numerically without assuming any relation between $m_D$ and $T$, we get a positive $\kappa_E$, as shown in Fig.~\ref{fig:delta_kappa}. This procedure to estimate $\kappa_E$ is often used in the literature, and we  denote the corresponding estimate as $\kappa_E^{\rm kin}$. 

Unless $m_D \ll T$, the approximation $\Pi_{00}(p,T) \simeq m_D^2$ is a rather poor one. We can improve it by considering the leading momentum dependence of the self-energy \cite{Brambilla:2020siz}:
\be
	\Pi_{00}(p,T)\simeq m_D^2-\frac{N_c}{4}g^2 T p\,.
	\label{eq:gluon_self_energy}
\ee
The leading momentum dependent term is quite large numerically, and also gauge parameter independent. Therefore, it makes sense to use the above expression for $\Pi_{00}(p,T)$ when evaluating the integral in Eq.~\eqref{eq:kappa_kinetic_integral}. We call the corresponding result resummed LO, and  denote it as $\bar \kappa_E^{\rm LO}$ in Fig.~\ref{fig:delta_kappa}. It turns out that $\bar \kappa_E^{\rm LO}$ is significantly larger than $\kappa_E^{\rm LO}$ and $\kappa_E^{\rm kin}$. 

The heavy-quark momentum diffusion coefficient was also calculated at NLO \cite{Caron-Huot:2007rwy,Caron-Huot:2008dyw}:
\be
	\kappa^{\rm NLO}_E = \kappa_E^{\rm LO} +\Delta \kappa_E^{\rm NLO},~\Delta \kappa_E^{\rm NLO}= \frac{g^4C_F T^2}{18\pi}\times 2.3302 N_c m_D\,.
	\label{eq:kappa_NLO}
\ee
Though formally it is an $O(g^5)$ effect, the NLO correction proves numerically substantial, rendering $\kappa_E^{\NLO}$ positive for all temperatures considered in this study, as one can see from Fig.~\ref{fig:delta_kappa}. We can also consider a resummed NLO expression for $\kappa_E$ by combining $\bar \kappa_E^{\rm LO}$ and $\Delta \kappa_E^{\rm NLO}$. Since the resummed LO result, $\bar \kappa_E^{\rm LO}$, contains part of the NLO correction, we have to subtract this to avoid double counting. To do so, we expand the part of the denominator related to the gluon propagator in Eq.~\eqref{eq:kappa_kinetic_integral} to LO in the momentum dependent self energy:
\be
	\frac{1}{(p^2+\Pi_{00}(p,T))^2}\simeq \frac{1}{(p^2+m_D^2)^2}+2\left(\frac{N_c}{4}g^2Tp\right)\times\frac{1}{(p^2+m_D^2)^3}+\dots
	\label{eq:p_integrand_expand}
\ee
We denote by $\delta_2$ the correction obtained by integrating Eq.~\eqref{eq:kappa_kinetic_integral} with only the second term in Eq.~\eqref{eq:p_integrand_expand}. Notice that for $m_D/T\ll 1$, we have $\delta_2 \sim g^5$. Then, the resummed NLO result for $\kappa_E$ can be written as
\be
    \bar \kappa_E^{\rm NLO} = \bar\kappa_E^{\rm LO} +\Delta \kappa_E^{\rm NLO}-\delta_2\,.
    \label{eq:kappa_E_delta}
\ee
This result is also shown in Fig.~\ref{fig:delta_kappa}, and is slightly smaller than the NLO result $\kappa^{\rm NLO}_E$. From Fig.~\ref{fig:delta_kappa} we see that, while there is still a significant difference between $\bar \kappa_E^{\rm NLO}$ and $\bar \kappa_E^{\rm LO}$, this difference is smaller than the difference between $\kappa_E^{\rm NLO}$ and $\kappa_E^{\rm LO}$.

Based on the above discussion, we can conclude that the best perturbative estimate for $\rho_E$ at small $\omega$ would be $\rho_E^{\rm{IR}}$ given by Eq.\eqref{eq:ir_rho} with $\kappa_E=\kappa_E^{\NLO}$. As we have seen in the previous section, the NLO result for $\omega\gg T$, labeled $\rho_E^{\rm{UV},\NLO}$ in Eq.~\eqref{eq:rho_NLO_E}, can account for the behavior of $G_E$ at small $\tau$. Therefore, a plausible perturbative form of the spectral function would be one that interpolates between these two regimes, i.e.
\be
    \rho_E^{\rm{pert}}(\omega,T)={\rm max}\left[\rho_E^{\rm{IR}}(\omega,T),\rho_E^{\rm{UV},\NLO}(\omega)\right]\,.
    \label{eq:rho_E_pert}
\ee
We denote by $G_E^{\rm{pert}}$ the correlation function obtained from this spectral function and evaluated at $\mu=2 \pi T$. In Fig.~\ref{fig:NLO_kappa}, we compare the chromo-electric correlator obtained on the lattice at the highest three temperatures with $G_E^{\rm{pert}}$. We see that the perturbative result can describe quite well the temperature and $\tau$ dependence of $G_E$ obtained on the lattice. The small discrepancy between the perturbative result and the lattice result seen at $T=500$ MeV disappears as the temperature increases. We note that at $T=10$ GeV, which is the highest temperature considered in this study, the ratio $G_E/G_E^{\NLOp}$ is consistent with one within one sigma (see Figs.~\ref{fig:con_flow_correlator_LOp}–\ref{fig:NLO_kappa}). Consequently, this marks an upper limit of sensitivity for extracting the diffusion coefficient from a lattice QCD approach.

\subsection{Chromo-magnetic correlators and spectral functions at NLO}
\label{subsec:kappa_B_PT}

\begin{figure}[t]
    \centerline{
        \includegraphics[width=0.5\textwidth]
        {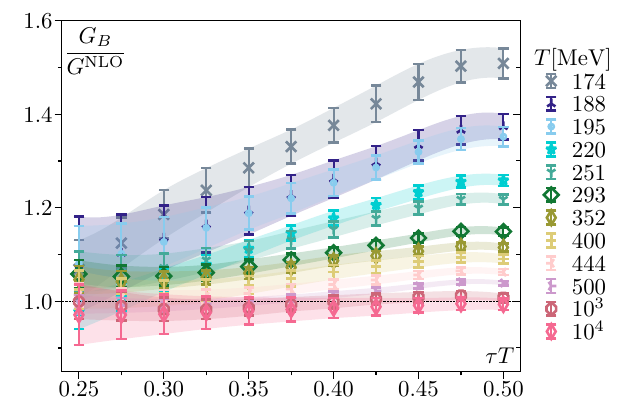}
    }
    \caption{Lattice results for $G_B$ normalized by the NLO result for $\mu=4$ GeV. For $T=10$ GeV, we show the results for $\mu=50$ GeV.}
    \label{fig:compB}
\end{figure}

The chromo-magnetic correlator has also been calculated at NLO \cite{Bouttefeux:2020ycy,Banerjee:2022uge,Altenkort:2023eav,delaCruz:2024cix}. The corresponding spectral function at $\omega \ll T$ has the form:
\be
	\rho^{\NLO}_{B}(\omega,T) \simeq \rho^{\rm{IR}}_{B}(\omega,T)=\frac{\kappa_{B}(T)}{2T}\omega\,,
	\label{eq:ir_rhoB}
\ee
where  \cite{Bouttefeux:2020ycy}
\begin{eqnarray}
    \kappa_B&=&\kappa_B^{\LO}=\frac{g^4 C_F T^3}{18\pi}\left( N_c \left[\ln \frac{2 T}{m_M}+1-\gamma_E+\frac{\zeta'(2)}{\zeta(2)} \right] \right.
    \nonumber\\
    &&\hspace{2.6cm}\left.+\frac{N_f}{2} \left[ \ln \frac{4 T}{m_M}+\frac{3}{2} -\gamma_E+\frac{\zeta'(2)}{\zeta(2)} \right ] \right)+{\cal O}(g^5)\,.
    \label{eq:kappa_B_LO}
\end{eqnarray}
However, unlike in the chromo-electric case, $\kappa_B$ cannot be calculated perturbatively. This is because $\kappa_B$ depends on the non-perturbative chromo-magnetic screening mass $m_M \sim g^2 T$ \cite{Bouttefeux:2020ycy}. This is the same Linde problem \cite{Linde:1980ts,Gross:1980br} that appears in the pressure at 4 loops and in the Debye mass calculations at 2 loops (NLO) \cite{Rebhan:1994mx,Braaten:1994pk}. Here it appears already at leading order because the correlation function involves soft chromo-magnetic fields at leading order, while for the pressure and the Debye mass soft chromo-magnetic fields first contribute at higher loop orders.
Thus, perturbation theory only gives the leading-log result for
$\kappa_B$.  This has been carefully discussed in Ref.~\cite{Bouttefeux:2020ycy}.

For $\omega \gg T$, the chromo-magnetic spectral function has the form \cite{Bouttefeux:2020ycy,Banerjee:2022uge,Altenkort:2023eav,delaCruz:2024cix}:
\begin{eqnarray}
	\rho_{B}^{\rm{UV},\NLO}(\omega,T,\mu)&=&\frac{g^2_{\MS}(\mu) C_F}{6 \pi}\omega^3\left\{1+\frac{g^2_{\MS}(\mu)}{(4\pi)^2}\left[-N_f\left(\frac{2}{3}\ln\frac{\mu^2}{4\omega^2}+\frac{26}{9}\right)\right.\right.+
	\label{eq:rho_NLO_B}\\
	&&\left.\left.N_c\left(\frac{5}{3}\ln\frac{\mu^2}{4\omega^2}+\frac{134}{9}-\frac{2\pi^2}{3}\right)\right]\right\}\,.
    \nonumber
\end{eqnarray}
Unlike $\rho_E$, it depends on the renormalization scale $\mu$ already at this order. Note that there was a mistake in the original result \cite{Banerjee:2022uge}, which was corrected in \cite{Altenkort:2023eav,delaCruz:2024cix}. This is a different mistake than the one found in the chromo-electric spectral function, even though both come from the same source. We can reconstruct $G_B^{\NLO}$ from the above spectral function. In Fig.~\ref{fig:compB}, we show the lattice results for $G_B$ normalized by $G_B^{\NLO}$ for $\mu=4$ GeV. As one can see from the figure, the NLO result can capture the behavior of $G_B$ at small $\tau$, at least for sufficiently high temperatures.


\section{Determination of the heavy quark diffusion coefficient}
\label{sec:spectral}

In this section, we will discuss the determination of the heavy-quark diffusion coefficient. The continuum and flow time extrapolated results on $G_E(\tau,T)$ and $G_B(\tau,T,\mu)$, together with the spectral representation of these correlators given by Eq.~\eqref{eq:spectral_rep}, allow for the determination of the transport coefficients $\kappa_E$ and $\kappa_B$. From these, one can determine the momentum and spatial diffusion coefficients for charm and bottom quarks, once we estimate their mean squared thermal velocity and squared momentum.

Since we only know the chromo-electric and chromo-magnetic correlator at discrete sets of $\tau$ values, the direct inversion of Eq.~\eqref{eq:spectral_rep} is not possible. Therefore, we need parametrizations of the spectral functions, $\rho_E(\omega,T)$ and $\rho_B(\omega,T)$, that encode their known behavior at small and large $\omega$. Fitting the lattice results on $G_{E,B}$ with such forms of the spectral function allows us to estimate transport coefficients, $\kappa_E$ and $\kappa_B$. In the following subsections, we will discuss the procedure for their determination separately. Then, we will discuss the determination of the average heavy-quark thermal squared velocity and squared momentum, and present the final results for the momentum and spatial diffusion coefficients.

\subsection{Determination of \texorpdfstring{$\kappa_E$}{}}
\label{subsec:spectral_E}

\begin{figure}[t]
    \centerline{
        \includegraphics[width=0.4\textwidth]{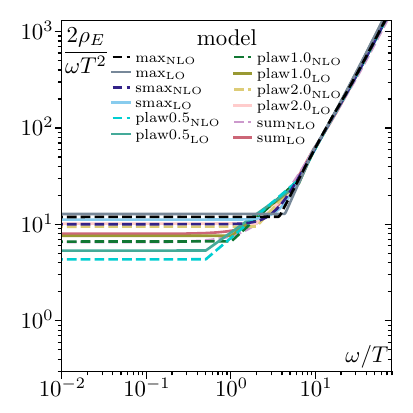}
        \includegraphics[width=0.4\textwidth]{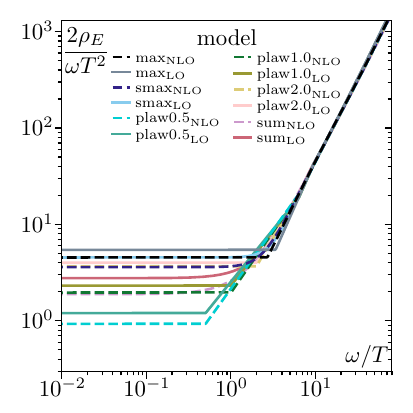}
    }
    \caption{Chromo-electric spectral function, reconstructed from the different fitting techniques described in the text, for two different temperature values. Left: $195$ MeV. Right: $400$ MeV. }
    \label{fig:spectral_fucntions}
\end{figure}

\begin{figure}[t]
    \centerline{
        \includegraphics[width=0.4\textwidth]{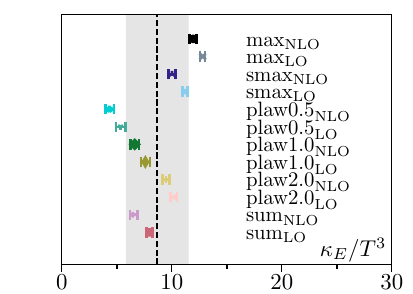}
        \includegraphics[width=0.4\textwidth]{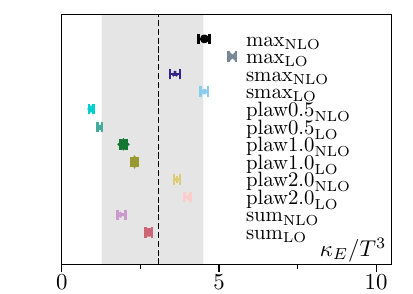}
    }
    \caption{Chromo-electric diffusion coefficient $\kappa_E/T^3$, extracted from the different fits described in the text, for two different temperature values. Left: $195$ MeV. Right: $400$ MeV. The gray bands denote the 68\% confidence intervals of the distributions of $\kappa_E$, see text. }
    \label{fig:kappa_E_over_T3}
\end{figure}

\begin{figure}
    \centerline{
        \includegraphics[width=0.4\textwidth]{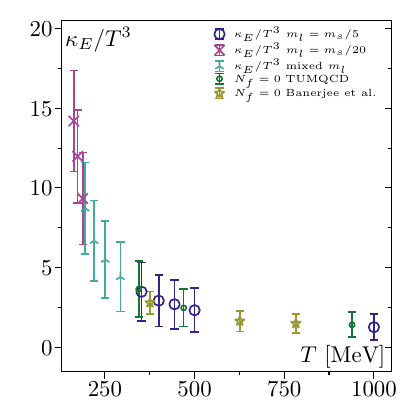}
        \includegraphics[width=0.4\textwidth]{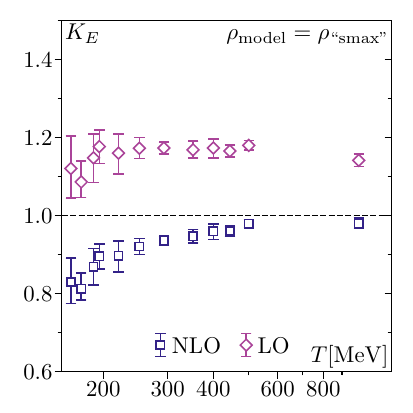}
    }
    \caption{Left: Temperature dependence of $\kappa_E/T^3$ obtained from
    the combined analysis of different fits, see text. We also show the values of $\kappa_E$ obtained in quenched QCD \cite{Brambilla:2020siz,Banerjee:2022gen}. Right: rescaling parameter $K_E$ obtained at different temperatures for the smax Ansatz.}
    \label{fig:fit_res_E}
\end{figure} 

For small $\tau$, the NLO result with running coupling constant $G_E^{\rm NLO+}$ describes the lattice results for the chromo-electric correlator $G_E$, c.f. Fig.~\ref{fig:con_flow_correlator_LOp}. Therefore, the NLO perturbative result for $\rho_E$ is a good approximation of the spectral function at larger $\omega$. Furthermore, we have seen that combining $\rho_E^{\rm UV, NLO}(\omega)$  and  $\rho_E^{\rm IR}(\omega,T)$, with the diffusion coefficient fixed to $\kappa_E^{\rm NLO}$ as described in \eqref{eq:rho_E_pert}, results in a spectral function that can describe fairly well the chromo-electric correlator at our highest three temperatures, c.f.\ Fig.~\ref{fig:NLO_kappa}. Thus, our Ansatz for $\rho_E(\omega,T)$ can be some combination of $\rho_E^{\rm IR}(\omega)$ and $\rho_E^{\rm UV, NLO}(\omega)$ or $\rho_E^{\rm LO}(\omega)$. We use the latter to check the sensitivity of $\kappa_E$ to the precise form of the perturbative determination of the spectral function at large $\omega$. To fully account for missing higher order perturbative corrections, we use $\rho_E^{\rm UV}(\omega)=K_E \rho_E^{\rm UV, NLO(LO)}(\omega)$, with $K_E$ a rescaling factor treated as an extra fit parameter. As in previous works
\cite{Brambilla:2020siz,Altenkort:2023oms,Altenkort:2023eav,Altenkort:2024spl}, we will use the following forms:

\begin{itemize}
\item Max Ansatz:
\be
	\rho_E^{\mathrm{max}}(\omega,T)=\max\{\rho_E^{\rm{IR}}(\omega,T),\rho_E^{\rm{UV}}(\omega)\}\,.
	\label{eq:rho_max}
\ee
\item Smooth max (smax) Ansatz:
\be
	\rho_E^{\mathrm{smax}}(\omega,T)=\sqrt{\left(\rho_E^{\rm{IR}}\left(\omega,T\right)\right)^2+\left(\rho_E^{\rm{UV}}\left(\omega\right)\right)^2}.
	\label{eq:rho_smax}
\ee
\item Power law Ansatz:
\be
	\rho_E^{\rm{plaw}}(\omega,T)=
	\left\{
	\begin{array}{cc}
		\rho^{\rm{IR}}(\omega,T) & 10^{-6}\leq\omega/T\leq \omega_{\rm{IR}} \\
		c\omega^p & \hspace{0.4cm} \omega_{\rm{IR}}\leq\omega/T\leq \omega_{\rm{UV}} \\
		\rho^{\rm{UV}}(\omega) & \hspace{0.1cm} \omega_{\rm{UV}} \leq\omega/T\leq 10^{3}
	\end{array} 
	\right.\,,
	\label{eq:rho_plaw}
\ee
\end{itemize}

The power law Ansatz has two additional parameters, $\omega_{\rm{IR}}$ and $\omega_{\rm{UV}}$.  Following \cite{Altenkort:2023oms}, and based on known physics considerations, we use $\omega_{\rm{IR}}/T=$0.5, 1.0, 2.0 and $\omega_{\rm{UV}}/T=2\pi$. Furthermore, the parameters $c$ and $p$ in this Ansatz are fixed by requiring continuity of the spectral function at $\omega=\omega_{\rm{IR}}$ and $\omega=\omega_{\rm UV}$. The max Ansatz and the power-law Ansatz do not correspond to smooth spectral functions. This is not a problem, since the fits to the lattice results are only sensitive to the integrals of the spectral function. Furthermore,  the non-smoothness of the above Ans\"atze can be fixed by using a smoothed theta function. For example, the max Ansatz can be written as:
\be
	\rho_E^{\rm{tanh}}(\omega,T)=\rho_E^{\rm{IR}}(\omega,T)\times\frac{1-\tanh[r_s(\omega-\omega_s)]}{2}+\rho_E^{\rm{UV}}(\omega)\times\frac{1+\tanh[r_s(\omega-\omega_s)]}{2}\,.
	\label{eq:rho_tanh}
\ee
For large value of $r_s$, the above equation closely approximates the max Ansatz. We also checked that using this form, with a large enough $r_s$, leads to the same value of $\kappa_E$ as the max Ansatz when used to fit the lattice results. In this work, we also consider a sum Ansatz:
\be
	\rho_E^{\mathrm{sum}}(\omega,T)=\rho_E^{\rm{IR}}(\omega,T)+\rho_E^{\rm{UV}}(\omega)\,.
	\label{eq:rho_sum}
\ee

Altogether, we have 12 different Ans\"atze to fit the lattice results for $G_E$, as illustrated in Fig.~\ref{fig:spectral_fucntions} for two representative temperatures, $T=195$ MeV and $T=400$ MeV. The corresponding fits work well as discussed in the Appendix~\ref{appendix:fits}. As an example, Fig.~\ref{fig:kappa_E_over_T3} displays the values of $\kappa_E$ extracted from the various fit forms at the same two representative temperatures. To determine the final estimate of $\kappa_E$, we generate $1000$ bootstrap samples for each of the 12 Ans\"atze and compute the mean and the 68\% confidence interval from the combined distribution. The resulting temperature dependence of $\kappa_E/T^3$ is shown in Fig.~\ref{fig:fit_res_E} (left), where we clearly see a decreasing trend with temperature, as expected. The final values are also listed in Table~\ref{tab:kappa_D_final}. For comparison, Fig.~\ref{fig:fit_res_E} includes the quenched ($N_f=0$) results from Refs.~\cite{Brambilla:2020siz,Banerjee:2022gen}, converted to physical units using $r_0T_c=0.7457$ for the deconfinement phase transition temperature \cite{Francis:2015lha} and $r_0=0.472$ fm. Remarkably, the (2+1)-flavor QCD results and the quenched results for $\kappa_E/T^3$ are consistent within errors.

In Fig.~\ref{fig:fit_res_E} (right), we also show the extracted values of the rescaling factor $K_E$ as a function of temperature for the smax Ansatz. Fits using the LO form of $\rho_E^{\rm UV}$ yield $K_E$ values that are systematically larger than one by about 20\%, while fits using the NLO form gives $K_E$ values that approach one at high temperatures. This behavior is expected: at high temperatures, the coupling $g^2$ becomes small, making the NLO prediction more accurate and reducing higher-order corrections. In contrast, the LO form lacks sufficient accuracy, even at the smallest $\tau$ probed by our lattice calculations. Although Fig.~\ref{fig:fit_res_E} shows results for $K_E$ only for the smax model, the qualitative behavior is similar across all fitting techniques.

\subsection{Determination of \texorpdfstring{$\kappa_B$}{}}
\label{subsec:spectral_B}

\begin{figure}[t]
    \centerline{
        \includegraphics[width=0.4\textwidth]{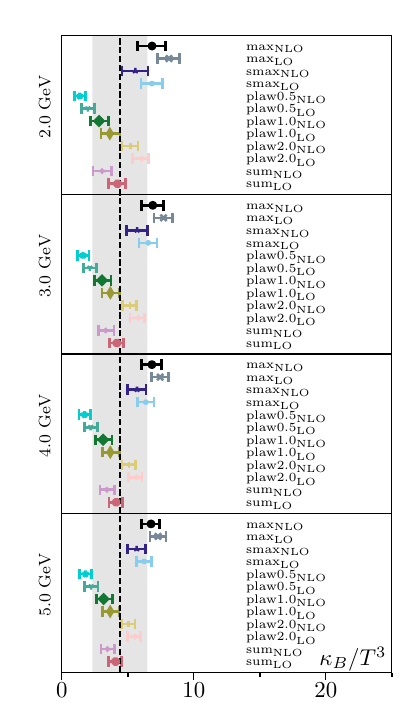}
        \includegraphics[width=0.4\textwidth]{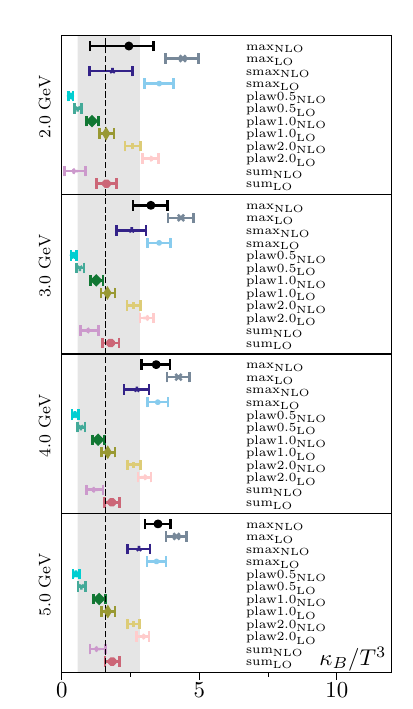}
    }
    \caption{Chromo-magnetic diffusion coefficient $\kappa_B/T^3$ for the different fitting techniques described in the text, for two different temperature values. Left: $195$ MeV. Right: $400$ MeV. The gray bands denote 68\% confidence intervals extracted from the distributions. }
    \label{fig:kappa_B_over_T3}
\end{figure}

\begin{figure}
    \centerline{
        \includegraphics[width=0.4\textwidth]{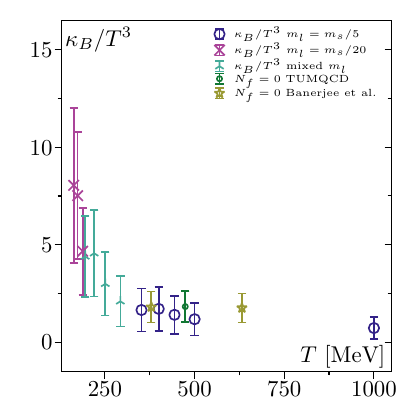}
        \includegraphics[width=0.4\textwidth]{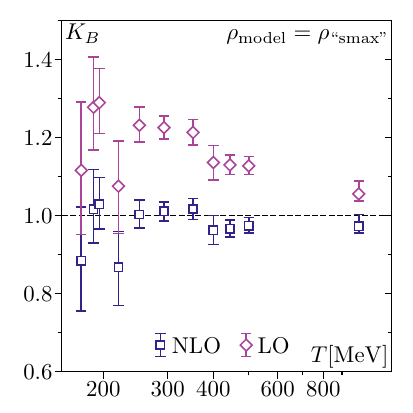}
    }
    \caption{Left: Temperature dependence of $\kappa_B/T^3$ obtained from
    the combined analysis of different fits, see text. Also shown there are the values of $\kappa_B$ obtained in quenched QCD \cite{Brambilla:2022xbd,Banerjee:2022uge} Right: $K_B$ obtained at different temperatures 
    for smax Ansatz. We do not display the results for $T=163$ MeV due to its large errors. 
    }
    \label{fig:fit_res_B}
\end{figure} 

Our procedure for determining $\kappa_B$ closely follows the one used for the determination of $\kappa_E$ as described above. Specifically, we consider 12 different Ans\"atze for the spectral function $\rho_B(\omega,T)$ that interpolate between the infrared form $\rho_B^{\rm IR}(\omega,T)=\kappa_B(T) \omega/(2 T)$  and the ultraviolet one $\rho_B^{\rm UV}=K_B \rho_B^{\rm UV,NLO (LO)}(\omega)$, i.e., use the fit forms given by Eqs.~\eqref{eq:rho_max}-\eqref{eq:rho_sum}, but with the subscript $E$ being replaced by the subscript $B$. Both $\kappa_B$ and $K_B$ are treated as fit parameters. 

As we stated in section~\ref{subsubsec:matching}, we do not use the physical scheme for $G_B$ introduced in \cite{Laine:2021uzs,Bouttefeux:2020ycy} and choose to leave the entire calculation in the $\MS$ scheme. However, an important step remains to be addressed. The chromo-magnetic correlator $G_B(\tau)$ in the $\MS$ scheme depends explicitly on the renormalization scale $\mu$, and as a result, the corresponding spectral function $\rho_B(\omega)$ inherits this scheme and scale dependence across all frequencies $\omega$. This dependence arises because the insertion of a magnetic $B_i$ field on a temporal Wilson line induces a non-trivial anomalous dimension, making any correlation function constructed from such operators a scheme and scale dependent quantity. 

The physically meaningful quantity is not the $\MS$ renormalized $\kappa_B^{\MS}$, but rather its renormalization-group invariant counterpart $\kappa_B^{\rm{phys}}$, which is defined through the zero-frequency limit of a heavy-quark current-current correlator. The relation between both the physical and $\MS$ correlators was worked out in \cite{Laine:2021uzs,Bouttefeux:2020ycy} and takes the form of the following multiplicative factor:
\be
    G^{\rm{phys}}_B(\tau T)=G^{\MS}(\tau T,\mu)\left[1+g^2\gamma_0\left(\ln\frac{\mu^2}{(4\pi T)^2}-2+2\gamma_{\rm E}\right)\right]\,.
    \label{eq:ms-to-physical}
\ee
In this work, we apply this correction as a final multiplicative rescaling to the $\kappa_B^{\MS}$ values extracted from the $\MS$ renormalized correlator, yielding the final result for $\kappa_B^{\rm{phys}}$, which we will simply denote as $\kappa_B$ in the following for brevity. This step removes the residual scale dependence, and restores the renormalization-group invariance up to higher-order corrections. We emphasize that this correction only applies to $\kappa_B$, as the chromo-electric correlator does not carry a non-trivial anomalous dimension in this setup.


To control the scale dependence, we perform the analysis using four values of the renormalization scale: $\mu=2,~3,~4$ and 5 GeV, except for $T=10$ GeV, where we use a higher scale of $\mu=50$ GeV. These choices are made to avoid large logarithmic corrections in the NLO expression for $\rho_B$ of Eq.~\eqref{eq:rho_NLO_B}. Ideally, the extracted value of $\kappa_B$ should be independent of $\mu$. In Fig.~\ref{fig:kappa_B_over_T3}, we show the results for $\kappa_B$ obtained from the different fits at each of the four values of $\mu$. From the figure it is clear that the dependence on $\mu$ is significantly smaller than both the statistical uncertainties and the split arising from the different fit forms. The fits themselves are also well behaved, as discussed in Appendix~\ref{appendix:fits}.

To obtain the final estimates for $\kappa_B$, we generate $1000$ bootstrap samples for each combination of fit Ansatz and renormalization scale. The resulting distributions are used to extract central values and uncertainties, where we report the median and the 68\% confidence interval as the final result. The temperature dependence of $\kappa_B/T^3$ is shown in Fig.~\ref{fig:fit_res_B} (left). We see that $\kappa_B/T^3$ is smaller than $\kappa_E/T^3$, and furthermore it decreases with increasing temperature, as expected. We also compare our results on $\kappa_B/T^3$ with the quenched results from Refs.~\cite{Brambilla:2022xbd,Banerjee:2022uge}, using the same procedure described above for converting $T/T_c$ to physical units. As for the electric diffusion coefficient, the quenched and 2+1-flavor results for $\kappa_B/T^3$ agree.

In the right panel of Fig.~\ref{fig:fit_res_B}, we show the values of $K_B$ extracted at various temperatures using the smax Ansatz. When the NLO form of $\rho_B^{\rm UV}$ is used, the fitted values of $K_B$ are consistent with one within errors. This indicates that the NLO expression provides a reliable description of the large-$\omega$ behavior of the chromo-magnetic spectral function.  Although we only present results for $K_B$ using the smax model, the qualitative behavior is similar across the other fitting forms.

\subsection{The temperature dependence of the heavy-quark diffusion coefficient}
\label{subsec:kappa_T_general}

In this subsection, we will discuss the temperature dependence of the momentum and spatial diffusion coefficient for charm and bottom quark. For this, we need to specify their mean squared thermal velocity and squared momentum. We will do this in terms of quasi-particle model. The need for some quasi-particle description is apparent in the Langevin description of the heavy-quark dynamics, which, in addition to $\kappa$ and the drag coefficient $\eta$, also depends on the heavy-quark mass. However, before discussing this, let us re-examine the temperature dependence of $\kappa_E$ and $\kappa_B$. 

\subsubsection{Temperature dependence of \texorpdfstring{$\kappa_{E,B}$}{}}
\label{subsubsec:kappa_T_E_B}

\begin{figure}[t]
    \centerline{
        \includegraphics[width=0.4\textwidth]{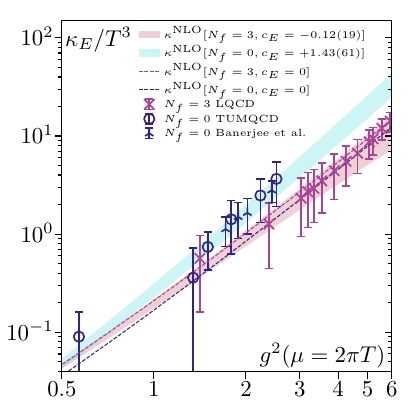}
    	\includegraphics[width=0.4\textwidth]{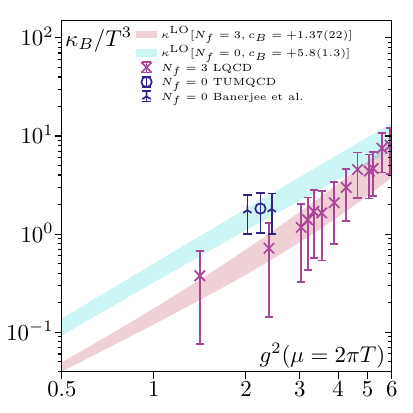}
    }
    \caption{Comparison of the coupling-strength dependence of the electric (left) and magnetic (right) diffusion coefficients. The coupling is evaluated at a scale $\mu=2\pi T$.  We also show the quenched $N_f=0$ results from the literature \cite{Brambilla:2020siz,Banerjee:2022gen,Brambilla:2022xbd,Banerjee:2022uge}. In the electric case (left plot), the black and red dashed lines are the NLO perturbative results for $N_f=0$ and $N_f=3$. The blue and pink bands are fits to the lattice data for $N_f=0$ and $N_f=3$, respectively, using Eqs.~\eqref{eq:kappaE_fit} and \eqref{eq:kappaB_fit} 
    }
\label{fig:g2_kappa}
\end{figure}

As noted above, $\kappa_E/T^3$ and $\kappa_B/T^3$ decrease with increasing temperature. At sufficiently high temperatures, we expect this dependence to be determined by the scale dependence of the coupling constant $g$. Therefore, we fix the renormalization scale $\mu$ to be $2 \pi T$ and recast the temperature dependence of $\kappa_{E,B}/T^3$ shown in the previous subsections as the dependence on $g^2(\mu=2 \pi T)$. This is shown in Fig.~\ref{fig:g2_kappa} using double logarithmic scale. It is easy to see that the $g$-dependence of $\kappa_{E,B}/T^3$ is compatible with $g^4$ and is incompatible with $g^2$ dependence. In this figure, we also show the quenched $N_f=0$ results from Refs.~\cite{Brambilla:2020siz,Banerjee:2022uge,Brambilla:2022xbd,Banerjee:2022gen}. For the $N_f=0$ we use $r_0 \LMS=0.647$ from FLAG 2024 Review \cite{FlavourLatticeAveragingGroupFLAG:2024oxs}, which is based on Refs.~\cite{Capitani:1998mq,Gockeler:2005rv,Brambilla:2010pp,Kitazawa:2016dsl,Ishikawa:2017xam,DallaBrida:2019wur,Bribian:2021cmg,Brambilla:2023fsi}, and $r_0=0.472$ fm. The NLO results for $\kappa_E$ with $\mu=2 \pi T$ are shown as dashed lines in the figure. The NLO result agrees with our lattice determination of $\kappa_E$ in 2+1-flavor QCD while remaining lower than the lattice results for quenched QCD.

Inspired by the perturbation theory and weak coupling calculations, we parametrize $\kappa_{E,B}/T^3$ as:
\begin{align}
    &\kappa_E(T)=\frac{g^4 C_F T^3}{18\pi}\left(\left[N_c+\frac{N_f}{2}\right]\left[\ln\frac{2T}{m_D}+\xi\right]+\frac{N_f\ln2}{2}+ 2.3302 N_c m_D+c_E g^2 \right)
    \label{eq:kappaE_fit}\,,\\
    &\kappa_B(T)=\frac{g^4 C_F T^3}{18\pi}\left( \left[N_c+\frac{N_f}{2}\right] \ln\frac{1}{g^2}+c_B \right)\,,
    \label{eq:kappaB_fit}
\end{align}
\textbf{using data corresponding to the large temperature region $T\leq 352$ MeV}. These parametrizations describe the lattice results on $\kappa_E$ and $\kappa_B$ very well, as can be seen
in Fig.~\ref{fig:g2_kappa}. From the fits using Eq.~\eqref{eq:kappaE_fit}, we get $c_E^{N_f=3}=-0.12(19)$ and $c_E^{N_f=0}=1.43(61)$,  Therefore, we see that the lattice results for $N_f=3$ are much closer to the perturbative NLO predictions than the $N_f=0$ lattice results. In the same way, from the fits to lattice results on $\kappa_B$ using Eq.~\eqref{eq:kappaB_fit}, we get $c_B=5.8(1.3)$ and $c_B=1.37(22)$ for $N_f=0$ and $N_f=3$, respectively.

\subsubsection{Quasi-particle model for charm and bottom degrees of freedom}
\label{subsubsec:QPM}

As previously discussed, Langevin dynamics requires specifying the masses of the charm and bottom quarks. At high temperatures, these heavy quarks are expected to be the relevant carriers of heavy-flavor quantum numbers. This expectation has been confirmed in quenched QCD through lattice calculations of susceptibilities associated with vector and axial-vector currents, as well as the scalar density, in the deconfined phase \cite{Petreczky:2008px}. This study showed that the susceptibilities can be well described by a quasi-particle model with temperature-dependent quark masses \cite{Petreczky:2008px}. In full QCD with 2+1 flavors, lattice calculations of generalized charm susceptibilities further support this picture, indicating that for $T>195$ MeV, the dominant degrees of freedom are indeed charm quarks \cite{Bazavov:2014yba,Mukherjee:2015mxc,Bazavov:2023xzm,Kaczmarek:2025dqt}. Based on this, we determine the charm quark mass, the mean squared thermal velocity $\vv$, and the mean squared thermal momentum $\pp$ using a quasi-particle description. This methodology follows the previous HotQCD study in Ref.~\cite{Altenkort:2023eav}.

In the quasi-particle model, the charm quark susceptibility can be written as:
\be
    \chi_c(T)=\frac{6}{\pi T} \int \frac{\mathrm{d}^3 p}{(2 \pi)^3} e^{-E_p/T},\ E_p=\sqrt{p^2+M_c^2(T)}\,,
    \label{eq:qmp_chi}
\ee
where $M_c(T)$ is the temperature dependent quasi-particle mass.
Similarly, for $\vv$ and $\pp$, we can write \cite{Petreczky:2008px}:
\begin{align}
    \vv &= \frac{6}{\pi T \chi_c(T)} \int \frac{\mathrm{d}^3 p}{(2 \pi)^3} \frac{p^2}{E_p^2}e^{-E_p/T}\,, 
    \label{eq:qmp_vsq}\\
    \pp &= \frac{6}{\pi T \chi_c(T)} \int \frac{\mathrm{d}^3 p}{(2 \pi)^3} p^2 e^{-E_p/T}\,. 
    \label{eq:qmp_psq}
\end{align}
In these equations, we have used Boltzmann distributions due to the large value of the charm-quark mass. To determine the temperature-dependent mass $M_c(T)$ for $T\ge 195$ MeV, we use the continuum-extrapolated lattice QCD results for the charm susceptibility $\chi_c(T)$ of the Wuppertal-Budapest Collaboration \cite{Bellwied:2015lba}. With these values of $M_c(T)$, we compute the mean squared thermal velocity $\vv$ and momentum $\pp$ using Eqs.~\eqref{eq:qmp_vsq} and \eqref{eq:qmp_psq}, respectively.

A detailed analysis of generalized charm susceptibilities shows that for $T<195$ MeV, charm hadron-like excitations, such as charm mesons and baryons, become important \cite{Mukherjee:2015mxc,Bazavov:2023xzm,Kaczmarek:2025dqt}. In this regime, the total charm pressure can be approximated as the sum of the partial pressures from charm mesons, charm baryons, and charm quarks \cite{Mukherjee:2015mxc,Bazavov:2023xzm,Kaczmarek:2025dqt}. Recent continuum-extrapolated results for these partial pressures have been estimated in Ref.~\cite{Kaczmarek:2025dqt}. Therefore, in the temperature range below $195$ MeV, it is more appropriate to estimate $\vv$ and $\pp$ using an extended quasi-particle model:
\begin{align}
    &\vv= \frac{1}{\pi T\chi_c(T)}\sum_i g_i\int \frac{\mathrm{d}^3 p}{(2 \pi)^3} \frac{p^2}{(E^i_p)^2}e^{-E_p^i/T}\,,\label{eq:v2_ext} \\
    &\pp = \frac{1}{\pi T\chi_c(T)}\sum_i g_i \int \frac{\mathrm{d}^3 p}{(2 \pi)^3} p^2 e^{-E_p^i/T}\,,
    \label{eq:p2_ext}
\end{align}
with $E_p^i=\sqrt{p^2+M_i^2}$. The sum goes over the different types of quasi-particle states: charm quarks, charm mesons and charm baryons. $M_i$ denotes the masses of the quasi-particles, and $g_i$ is the corresponding degeneracy factor (for quarks $g_q=6$). 

Furthermore, the charm susceptibility now is given by
\be
    \chi_c(T)=\frac{1}{\pi T}\sum_i g_i \int \frac{d^3 p}{(2 \pi)^3} e^{-E_p^i/T}\,.
    \label{eq:qmp_chi_ext}
\ee
It turns out that the values of the averaged thermal velocity and thermal momentum do not depend much on whether they are evaluated in the simple quasi-particle model, i.e. assuming only charm quarks, or the extended quasi-particle  model, as long as the calculations are constrained by the lattice QCD results on the charm susceptibilities.  This is discussed in the Appendix~\ref{appendix:qpm}, where we also give further details about the determination of $\vv$ and $\pp$ with the estimates of their uncertainties, see Tab.~\ref{tab:thermal_variables_charm} and Fig.~\ref{fig:v2_p2}.

Unfortunately, not much is known about the open bottom degrees of freedom in (2+1)-flavor QCD from lattice QCD calculations. Based on the heavy-quark symmetry arguments, we assume that bottom degrees of freedom behave similarly to the charm degrees of freedom. Namely, we assume that for $T \ge 195$ MeV, bottom quarks are the dominant degrees of freedom, while for lower temperature we also use the extended quasi-particle model that includes bottom quarks and open beauty mesons and baryons.  Due to absence of lattice QCD data, we rather use a hadron resonance gas (HRG) model for the bottom susceptibility to determine the temperature dependent quasi-particle bottom mass $M_b(T)$ along with the values of $\vv$ and $\pp$. Again, as discussed in the Appendix~\ref{appendix:qpm}, the values of $\vv$ and $\pp$ for bottom degrees of freedom are not very sensitive to our assumptions, cf. Tab.~\ref{tab:thermal_variables_bottom}.

\subsubsection{The temperature dependence \texorpdfstring{$\kappa$}{} and \texorpdfstring{$D_s$}{} for charm and bottom quarks}
\label{subsubsec:kappa_D}

\begin{figure}[t]
    \centerline{
        \includegraphics[width=0.4\textwidth]{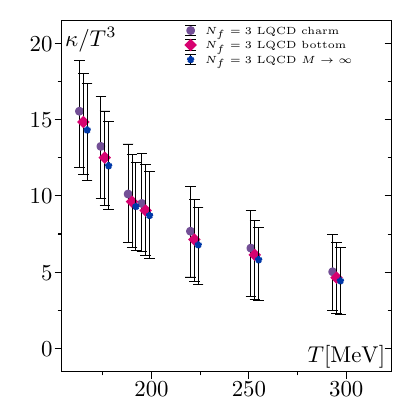}
        \includegraphics[width=0.4\textwidth]{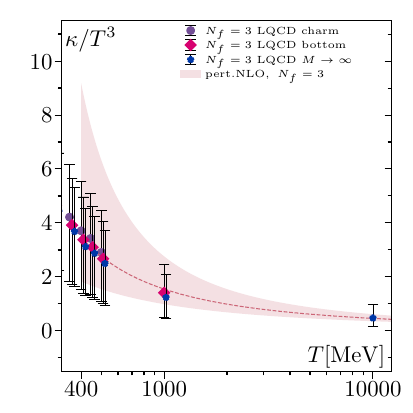}
    }
    \caption{Temperature dependence of the total momentum diffusion coefficient for charm, bottom and infinitely heavy quarks. For large temperatures (right plot) we compare the results with the expected NLO result of Eq.~\eqref{eq:kappa_NLO} for infinitely heavy quark. We use $\mu=2 \pi T$
    for the central value of the renormalization scale (dashed line) and vary the renormalization scale by factor of two around this central value. The corresponding variation of the NLO prediction is the width of the band.
    }
    \label{fig:kappa_total}
\end{figure}

Having determined the mean squared thermal velocity and momentum of charm and bottom quarks above the crossover temperature, we can now compute the total heavy-quark momentum and spatial diffusion coefficient. Where relevant, we will also compare these results to the infinite-mass limit. In Fig.~\ref{fig:kappa_total}, we present our final results for the total momentum diffusion coefficient $\kappa$, obtained using the thermal averages listed in Tabs.~\ref{tab:thermal_variables_charm} and \ref{tab:thermal_variables_bottom}.
The uncertainties in $\kappa$ arising from variations in the thermal averages (due to different quasi-particle models) are much smaller than the dominant uncertainties associated with the determination of $\kappa_E$ and $\kappa_B$. Moreover, the dependence of $\kappa$ on the heavy-quark mass is quite small up to the highest temperatures for which the inverse mass expansion remains valid.

\begin{figure}[t]
    \centerline{
        \includegraphics[width=0.4\textwidth]{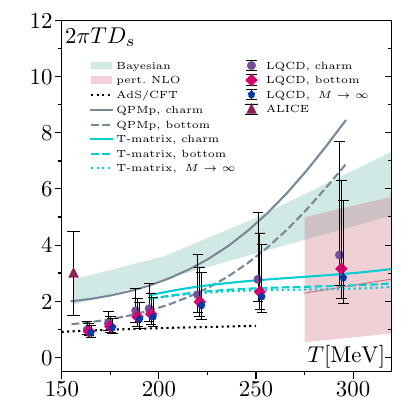}
        \includegraphics[width=0.4\textwidth]{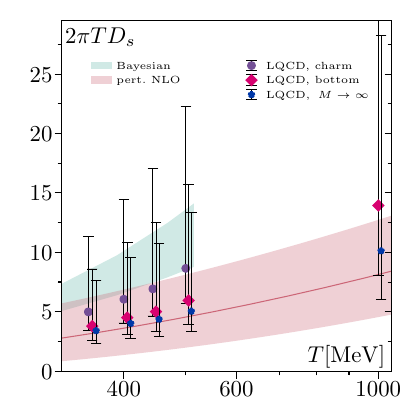}
    }
    \caption{Final results for the spatial diffusion coefficients $2\pi TD_s$, for the charm, bottom and infinitely-heavy quarks, compared with several other estimations and determinations. We include the QPMp estimates \cite{Sambataro:2024mkr}, the bayesian analysis of \cite{Xu:2017obm}, the data-driven result of the ALICE collaboration and the result extracted from T-matrix formalism \cite{Tang:2023tkm}. The dotted line represents the strongly coupled estimated based on AdS/CFT  \cite{Casalderrey-Solana:2006fio}. In addition, we also include the NLO perturbative QCD estimate \cite{Caron-Huot:2007rwy}, where the band corresponds to variations of the renormalization scale from $\mu = \pi T$ (upper boundary) to $\mu = 4\pi T$ (lower boundary). The central value of the perturbative band is obtained using a scale $\mu = 2\pi T$. We have split the figure into two parts: (right) one for the low temperature behavior and (left) a second one with the high temperature behavior, where we have only included the bottom and infinitely heavy quark for $T=1$ GeV.}
    \label{fig:2piTD}
\end{figure}

We obtain the spatial diffusion coefficient by combining the values of $\kappa$ with the results of $\pp$ given in Appendix~\ref{appendix:qpm} according to Eq.~\eqref{eq:D_s}. Fig.~\ref{fig:2piTD} shows the dependence of $D_s$ on the temperature. We have split the figure into two parts to show the low and high temperature dependence separately. The fact that the errors grow drastically with increasing temperature is due to the inverse relation between $\kappa$ and $D_s$, cf. Eq.~\eqref{eq:D_s}. This inverse relation also results in significantly asymmetric errors of $D_s$. 

For $195$ MeV $\le T \le$ 293 MeV, our results agree with the previous HotQCD results \cite{Altenkort:2023oms,Altenkort:2023eav}. Additionally, our results show a significant increase in $2\pi TD_s$ with increasing temperature, and are in agreement with the NLO perturbative predictions for a wide range of temperatures. At low temperatures, say $T<300$ MeV this agreement is rather accidental, as the perturbative result cannot be trusted at such low temperature. 

The heavy-quark diffusion coefficient has also been studied in the context of strongly coupled $N=4$ supersymmetric gauge theory using the AdS/CFT correspondence \cite{Casalderrey-Solana:2006fio,Herzog:2006gh}. These calculations give $D_s=2/(\pi T \sqrt{\lambda})$, where $\lambda=g^2 N_c$ is the ’t Hooft coupling. By setting $N_c=3$, and identifying $g^2$ with the QCD coupling $g^2_{N_f=3}(\mu=2\pi T)$, we obtain a strong-coupling estimate for $D_s$ based on AdS/CFT, shown as the black dashed line in Fig.~\ref{fig:2piTD}. Remarkably, our lattice results for $D_s$ at the four lowest temperatures are close to this prediction. In addition, we note that another strong-coupling approach in Ref.~\cite{Andreev:2017bvr} also yields $2 \pi T D_s \sim 1$ for $T<250$ MeV.

In Fig.~\ref{fig:2piTD}, we compare our result on $D_s$ with other phenomenological determinations of the heavy-quark diffusion coefficient. Our result is systematically lower than those derived from Bayesian analysis of HIC data \cite{Xu:2017obm}, and is also lower than the data-driven result of the ALICE collaboration at $T\sim 150$ MeV \cite{ALICE:2021rxa}. However, our results are in agreement with those obtained using the T-matrix formalism \cite{Tang:2023tkm,Tang:2024dkz}. In Fig.~\ref{fig:2piTD}, we also show the latest results from the Catania group \cite{Sambataro:2024mkr} obtained by using a refined version of their original quasi-particle description, known as QPMp. Their corresponding results are in agreement with our determination for bottom quark at low temperature, but remain much larger for charm quark and over-predict $D_s$ for both charm and bottom quarks for temperatures $T>250$ MeV.


\section{Conclusions}
\label{sec:conclusions}
In this paper, we present the first complete (2+1)-flavor lattice QCD calculation of the heavy-quark diffusion coefficient for a wide range of temperatures. For low temperatures, we extracted $\kappa$ at almost physical point, using a physical strange-quark mass $m_s$ and near-physical light-quark mass $m_l = m_s/20$, corresponding to a pion mass of $m_\pi \simeq 160$ MeV. On the other hand, for high temperatures we use a physical strange-quark mass $m_s$ and $m_l = m_s/5$, corresponding to a pion mass of $m_\pi \simeq 320$ MeV because light quark mass effects are not important in this region. This study extends previous calculations of the HotQCD collaboration \cite{Altenkort:2023oms,Altenkort:2023eav} to the physical point for low temperatures, extending the temperature range to 163 MeV $\le T \le$ 10000 MeV. 

Our analysis of the chromo-electric correlation function employs the correct high-energy behavior of the corresponding spectral function, computed in Refs.~\cite{delaCruz:2024cix, Scheihing-Hitschfeld:2023tuz}. Although this correction does not significantly change the extracted value of the heavy-quark diffusion coefficient within errors, it resolves the puzzle of the large rescaling factor $K_E$ observed in previous works \cite{Brambilla:2020siz, Altenkort:2023oms}. As discussed in Section~\ref{subsec:kappa_E_high_T}, this improvement shows, for the first time, a direct agreement between lattice data and the perturbative NLO prediction at the level of the correlation function. This, in turn, reinforces the validity of our spectral function reconstruction strategy.

Our results show that close the crossover temperature, the spatial heavy-quark diffusion coefficient approaches the strongly coupled AdS/CFT estimate~\cite{Casalderrey-Solana:2006fio,Herzog:2006gh}, supporting the picture of the quark-gluon plasma as a strongly-coupled system and an almost perfect fluid, in which heavy quarks rapidly thermalize~\cite{Rapp:2018qla,Dong:2019byy,He:2022ywp}. At high temperatures, our results are consistent within errors with the NLO perturbative QCD prediction~\cite{Caron-Huot:2007rwy}. Across the entire temperature range, the lattice QCD results remain systematically lower than the estimates from Bayesian reconstructions~\cite{Xu:2017obm} and data-driven extractions~\cite{ALICE:2012ab}, also showing a weaker dependence on the heavy-quark mass compared to certain phenomenological models.

Clearly, the results presented in this work provide first principle QCD benchmark for the heavy-quark transport and thus, will be very useful for the phenomenology of heavy flavor probes in heavy ion collisions.


\appendix
\vskip0.5truecm
\centerline{\Large \bf Appendices}
\section{Continuum extrapolations of \texorpdfstring{$G_E$}{} and \texorpdfstring{$G_B$}{}}
\label{appendix:extrapolation}

\begin{figure}
    \centerline{
        \includegraphics[width=0.4\textwidth]{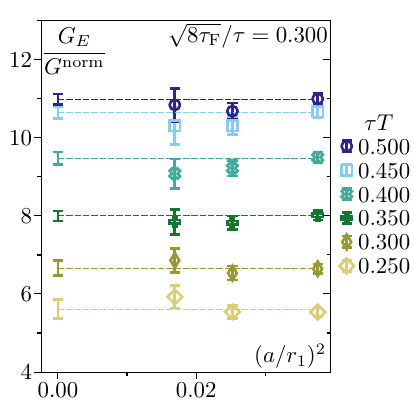}
        \includegraphics[width=0.4\textwidth]{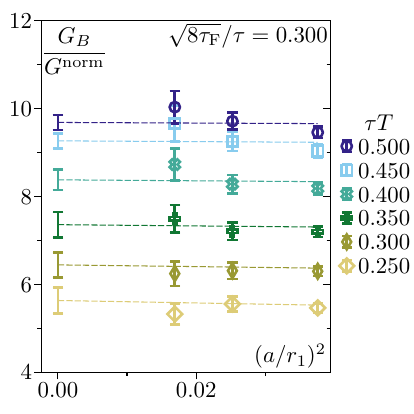}
    }
    \centerline{
       \includegraphics[width=0.4\textwidth]{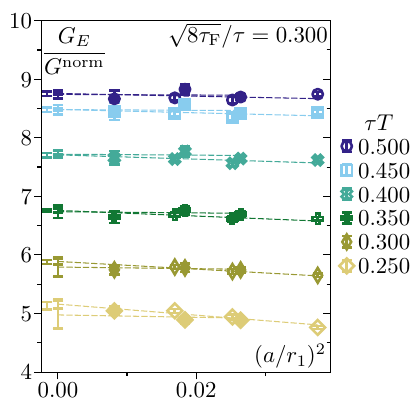}
       \includegraphics[width=0.4\textwidth]{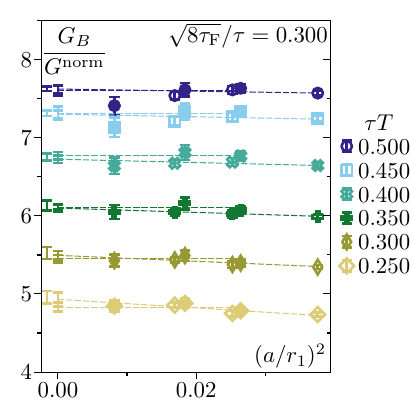}
    }
    \caption{Continuum extrapolations for $G_E$ (left) and $G_B$ (right) at relative flow time $\sqrt{8 \tau_F}/\tau=0.3$. The top row shows results for $T=163$ MeV, while the bottom row corresponds to $T=195$ MeV. Filled symbols represent data with $m_l=m_s/5$, while open symbols represent $m_l=m_s/20$. For $T=195$ MeV, the dashed lines indicate independent continuum extrapolations for each mass setup, while the points slightly to the left of $a/r_1=0$ show the combined continuum extrapolated results.}
    \label{fig:cont_extr_low}
\end{figure}

\begin{figure}
    \centerline{
        \includegraphics[width=0.4\textwidth]{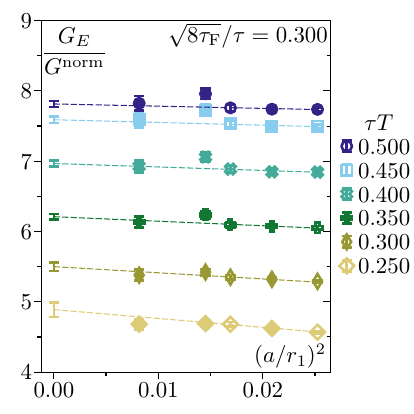}
        \includegraphics[width=0.4\textwidth]{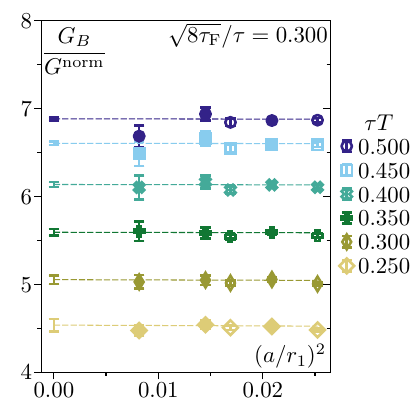}
    }
    \centerline{
       \includegraphics[width=0.4\textwidth]{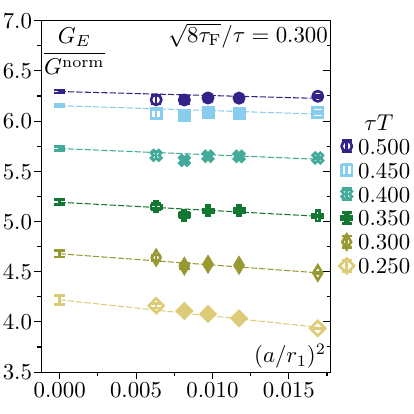}
       \includegraphics[width=0.4\textwidth]{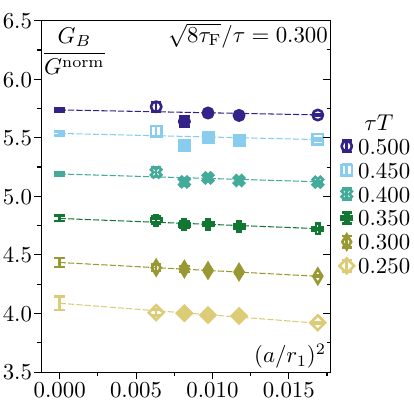}
    }
    \caption{Continuum extrapolations for $G_E$ (left) and $G_B$ (right) at relative flow time $\sqrt{8 \tau_F}/\tau=0.3$. The top row shows results for $T=220$ MeV, while the bottom row shows results for $T=293$ MeV. Filled symbols represent data with $m_l=m_s/5$, and open symbols represent $m_l=m_s/20$.}
    \label{fig:cont_extr_inter}
\end{figure}

\begin{figure}
    \centerline{
        \includegraphics[width=0.4\textwidth]{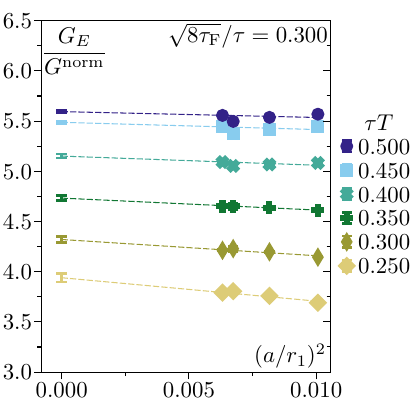}
        \includegraphics[width=0.4\textwidth]{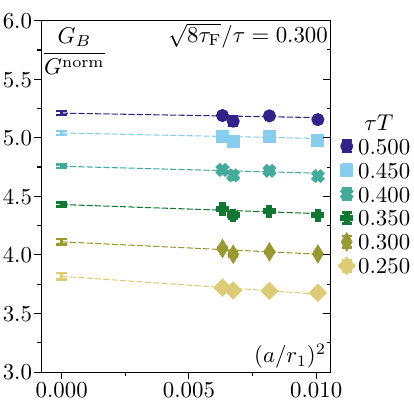}
    }
    \centerline{
       \includegraphics[width=0.4\textwidth]{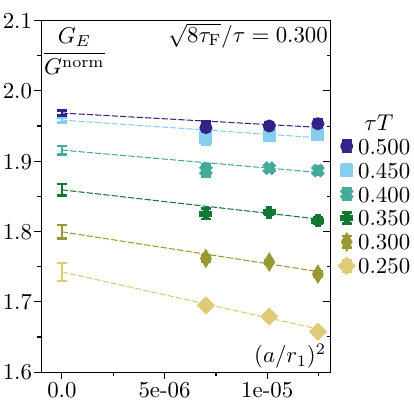}
       \includegraphics[width=0.4\textwidth]{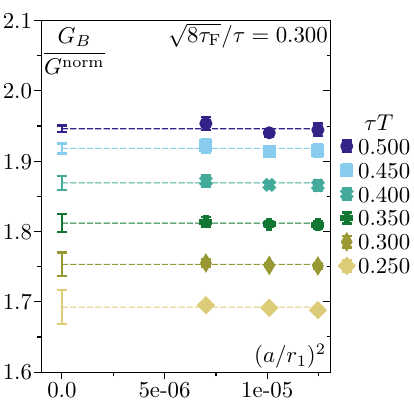}
    }
    \caption{Continuum extrapolations for $G_E$ (left) and $G_B$ (right) at relative flow time $\sqrt{8 \tau_F}/\tau=0.3$. The top row shows results for $T=352$ MeV, while the bottom row shows results for $T=10^4$ MeV. }
    \label{fig:cont_extr_high}
\end{figure}

In Figs.~\ref{fig:cont_extr_low}, \ref{fig:cont_extr_inter}, and \ref{fig:cont_extr_high}, we show representative continuum extrapolations of the chromo-electric and chromo-magnetic correlators in the low-, intermediate-, and high-temperature regions, respectively. In the low-temperature region $T \le 205$ MeV, the extrapolations are based on data with $m_l = m_s/20$, except at $T = 195$ MeV, where we perform separate extrapolations for $m_l = m_s/5$ and $m_l = m_s/20$, as well as a combined extrapolation. As shown in Fig.~\ref{fig:cont_extr_low}, the continuum results obtained from the two light quark masses, as well as their combination, agree within errors at $T = 195$ MeV. This indicates that the effects of the light quark mass are already small at this temperature. Consequently, in the intermediate-temperature region, we perform combined extrapolations using both $m_l = m_s/5$ and $m_l = m_s/20$ data, see Fig.\ref{fig:cont_extr_inter}. In the high-temperature region $T \ge 352$ MeV, we use only the $m_l = m_s/5$ data for the extrapolations, as shown in Fig.~\ref{fig:cont_extr_high}.

\section{Fits of the chromo-electric and chromo-magnetic correlators}
\label{appendix:fits}

\begin{figure}[t]
    \centerline{
        \includegraphics[width=0.4\textwidth]{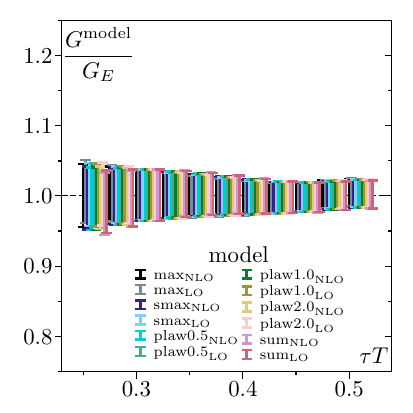}
        \includegraphics[width=0.4\textwidth]{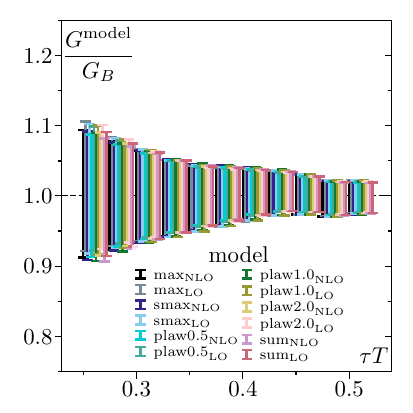}
    }
    \centerline{
        \includegraphics[width=0.4\textwidth]{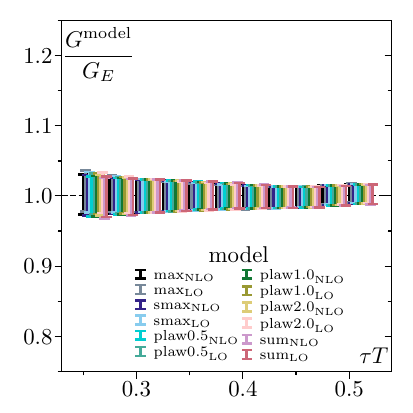}
        \includegraphics[width=0.4\textwidth]{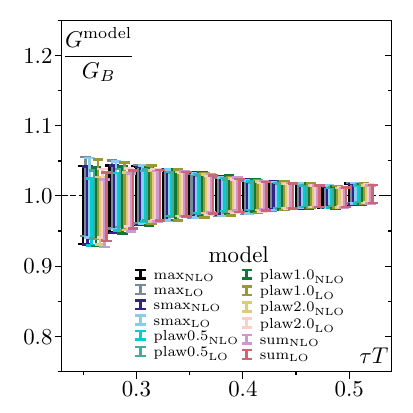}
    }
    \caption{Reconstructed correlators from the different fitting techniques described in the text, for 2 different temperature values. Upper left: electric correlator for $T=195$ MeV. Upper right: magnetic correlator for $T=195$ MeV at $\mu=4$ GeV. Lower left: electric correlator for $T=400$ MeV. Lower right: magnetic correlator for $T=400$ MeV at $\mu=4$ GeV.}
    \label{fig:fit_correlators}
\end{figure}

The spectral function Ans\"atze given in Eqs.~\eqref{eq:rho_max}, \eqref{eq:rho_smax}, \eqref{eq:rho_plaw}, and \eqref{eq:rho_sum} provide an excellent description of the continuum and flow-time extrapolated lattice results for both the chromo-electric and chromo-magnetic correlators. This is illustrated in Fig.~\ref{fig:fit_correlators}, where we show the ratio of the fitted correlators to the directly computed lattice correlators at two representative temperatures, $T=195$ MeV and $T=400$ MeV. In both cases, the ratio is very close to one across the entire range, confirming the quality of the fits.

\section{Quasi-particle model for charm and bottom degrees of freedom}
\label{appendix:qpm}

\begin{figure}[ht]
    \centerline{
        \includegraphics[width=0.4\textwidth]{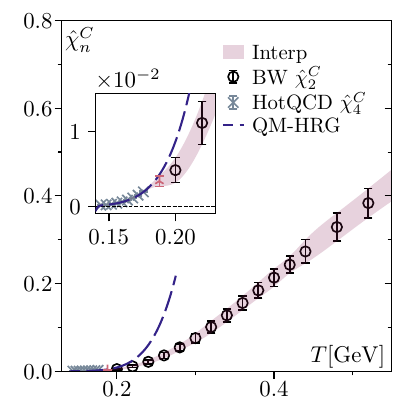}
        \includegraphics[width=0.4\textwidth]{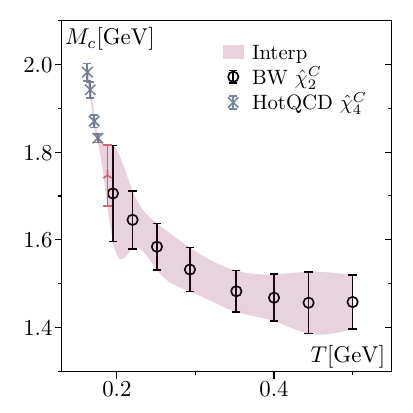}
    }
    \caption{Left: Lattice results together with the interpolation of the charm susceptibility $\hat \chi_2^C$ from the BW collaboration \cite{Bellwied:2015lba} with the recent $\hat \chi_4^C$ results from the HotQCD collaboration \cite{Kaczmarek:2025dqt}. For comparison, we also show the prediction from a full hadron resonance gas model QM-HRG. Right: interpolation of the charm-quark mass extracted from the susceptibility data of the BW and HotQCD collaborations. The orange point indicates the interpolated value at $T=188$ MeV.}
    \label{fig:chi-charm}
\end{figure}

\begin{figure}[ht!]
    \centerline{
        \includegraphics[width=0.4\textwidth]{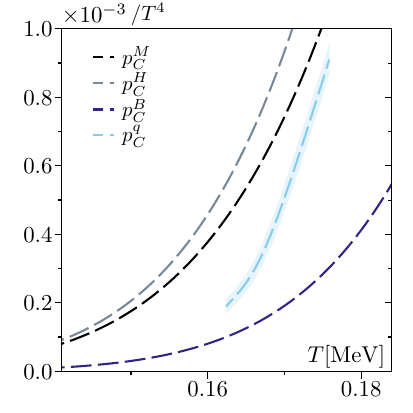}
        \includegraphics[width=0.4\textwidth]{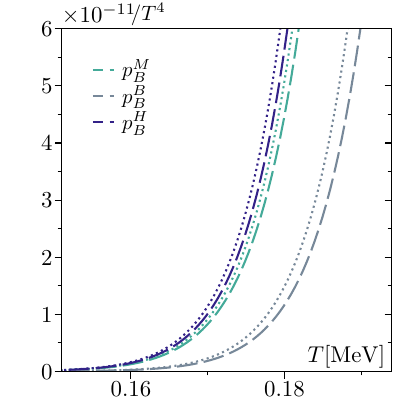}
    }
    \caption{Charm and bottom partial pressures in units of $T^4$. Left: charm quark partial pressure obtained by HotQCD determination \cite{Kaczmarek:2025dqt} compared to the charm meson, charm baryon and total charm hadron pressure from a 1S1P HRG model. Right: bottom partial pressures, showing results from the 1S1P HRG (dotted lines) and full HRG (dashed lines). These estimates are subsequently used to extract the bottom-quark mass as a function of temperature across the different models discussed in the text.
    }
    \label{fig:preassures}
\end{figure}

\begin{figure}[ht]
    \centerline{
        \includegraphics[width=0.4\textwidth]{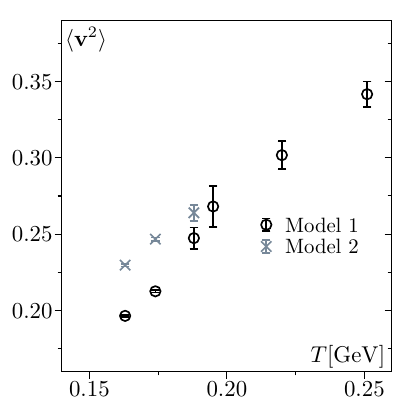}
        \includegraphics[width=0.4\textwidth]{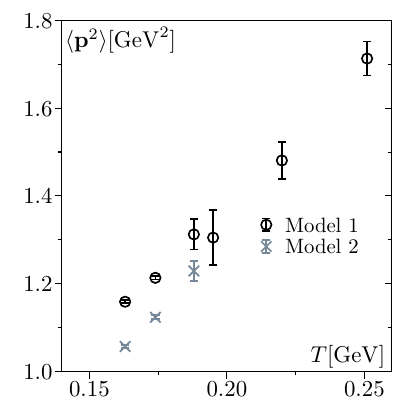}
    }
    \centerline{
        \includegraphics[width=0.4\textwidth]{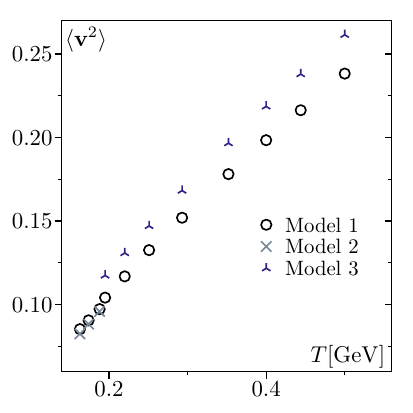}
        \includegraphics[width=0.4\textwidth]{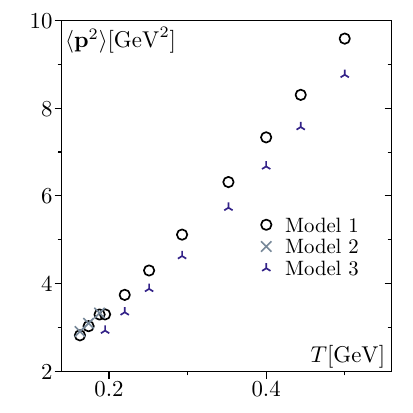}
    }
    \caption{Comparison of $\vv$ and $\pp$ for charm (top row) and bottom (bottom row) quarks, based on the different models described in the text. Model 1 employs a hadron resonance gas at low temperatures, whereas Model 2 assumes a free quark gas across the entire temperature range. For the bottom quark, we also present Model 3, where we fix the bottom-quark mass to 4.5 GeV for temperatures above $195$ MeV.
    }
    \label{fig:v2_p2_compare}
\end{figure}

\begin{table}[ht]
\centering
\begin{tabular}{c|ccc|ccc}
\toprule
$T$ [MeV]	& $M_{c_1}$	&   $\vv_{c_1}$		&	$\langle \mathbf{\hat{p}}^2\rangle_{c_1}$	&	$M_{c_2}$	&   $\vv_{c_2}$	& $\langle \mathbf{\hat{p}}^2\rangle_{c_2}$ \\
\midrule
163	 & 	 1.977(20)	 & 	 0.19633(70)	 & 	 1.199(13)	 & 	 1.7266(78)	 & 	 0.22962(84)	 & 	 1.2512(80)	 \\
174	 & 	 1.845(11)	 & 	 0.21244(77)	 & 	 1.2599(84)	 & 	 1.6863(85)	 & 	 0.2466(10)	 & 	 1.2760(89)	 \\
188	 & 	 1.747(70)	 & 	 0.2473(71)	     & 	 1.332(65)	 & 	 1.673(42)	 & 	 0.2639(52)	 & 	 1.302(42)	 \\
\midrule
195	 & 	 1.71(11)	 & 	 0.268(13)	 & 	 1.31(11)	 &   ------  &   ------  &   ------  \\
220	 & 	 1.645(66)	 & 	 0.3018(91)	 & 	 1.364(64)	 &   ------  &   ------  &   ------  \\
251	 & 	 1.584(53)	 & 	 0.3417(83)	 & 	 1.436(52)	 &   ------  &   ------  &   ------  \\
293	 & 	 1.532(50)	 & 	 0.3896(86)	 & 	 1.535(48)	 &   ------  &   ------  &   ------  \\
352	 & 	 1.482(47)	 & 	 0.4485(89)	 & 	 1.678(45)	 &   ------  &   ------  &   ------  \\
400	 & 	 1.468(53)	 & 	 0.488(10)	 & 	 1.789(50)	 &   ------  &   ------  &   ------  \\
444	 & 	 1.456(70)	 & 	 0.520(14)	 & 	 1.893(64)	 &   ------  &   ------  &   ------  \\
500	 & 	 1.458(62)	 & 	 0.555(12)	 & 	 2.018(54)	 &   ------  &   ------  &   ------  \\
\bottomrule
\end{tabular}
\caption{Mean squared thermal velocity $\vv$ and scaled mean squared thermal momentum $\langle \mathbf{\hat{p}}^2\rangle \equiv \pp/(3 M_c T)$ of charm degrees of freedom extracted from the quasi-particle models discussed in the text. Here, $M_c$ is the charm-quark mass in the corresponding quasi-particle model. All masses are given in GeV. }
\label{tab:thermal_variables_charm}
\end{table}

\begin{table}[ht!]
\centering
\begin{tabular}{c|ccc|ccc|ccc}
\toprule
$T$ [MeV]	&   $M_{b_1}$	&   $\vv_{b_1}$		&	$\langle \mathbf{\hat{p}}_{b_1}^2\rangle$	&	$M_{b_2}$	&   $\vv_{b_2}$		&	$\langle \mathbf{\hat{p}}_{b_2}^2\rangle$  &	$M_{b_3}$	&   $\vv_{b_3}$		&	$\langle \mathbf{\hat{p}}_{b_3}^2\rangle$ \\
\midrule
163	 & 	 5.136	 & 	 0.0852	 & 	 1.1227	 & 	 5.529	 & 	 0.0824	 & 	 1.0752	 &   ------  &   ------  &   ------  \\
174	 & 	 5.104	 & 	 0.0905	 & 	 1.1364	 & 	 5.482	 & 	 0.0883	 & 	 1.0811	 &   ------  &   ------  &   ------  \\
188	 & 	 5.059	 & 	 0.0972	 & 	 1.1547	 & 	 5.412	 & 	 0.0958	 & 	 1.0889	 &   ------  &   ------  &   ------  \\
\midrule
195	 & 	 5.059	 & 	 0.1041	 & 	 1.1141	 &   ------  &   ------  &   ------  & 	 4.500	 & 	 0.1174	 & 	 1.1117	 \\
220	 & 	 5.059	 & 	 0.1168	 & 	 1.1210	 &   ------  &   ------  &   ------  & 	 4.500	 & 	 0.1308	 & 	 1.1265	 \\
251	 & 	 5.059	 & 	 0.1325	 & 	 1.1285	 &   ------  &   ------  &   ------  & 	 4.500	 & 	 0.1470	 & 	 1.1450	 \\
293	 & 	 5.059	 & 	 0.1519	 & 	 1.1507	 &   ------  &   ------  &   ------  & 	 4.500	 & 	 0.1682	 & 	 1.1702	 \\
352	 & 	 5.059	 & 	 0.1781	 & 	 1.1824	 &   ------  &   ------  &   ------  & 	 4.500	 & 	 0.1966	 & 	 1.2062	 \\
400	 & 	 5.059	 & 	 0.1984	 & 	 1.2085	 &   ------  &   ------  &   ------  & 	 4.500	 & 	 0.2186	 & 	 1.2358	 \\
444	 & 	 5.059	 & 	 0.2163	 & 	 1.2326	 &   ------  &   ------  &   ------  & 	 4.500	 & 	 0.2380	 & 	 1.2632	 \\
500	 & 	 5.059	 & 	 0.2383	 & 	 1.2637	 &   ------  &   ------  &   ------  & 	 4.500	 & 	 0.2615	 & 	 1.2985	 \\
1000 & 	 5.059	 & 	 0.3982	 & 	 1.5547	 &   ------  &   ------  &   ------  & 	 4.500	 & 	 0.4299	 & 	 1.6304	 \\
\bottomrule
\end{tabular}
\caption{Mean squared thermal velocity $\vv$ and scaled mean squared thermal momentum $\langle \mathbf{\hat{p}}^2\rangle \equiv \pp/(3 M_b T)$ of bottom degrees of freedom extracted from the quasi-particle models discussed in the text. Here, $M_b$ is the bottom-quark mass in the corresponding quasi-particle model. All masses are given in GeV. }
\label{tab:thermal_variables_bottom}
\end{table}

\begin{table}[ht]
\centering
\begin{tabular}{c|cc|cc|cc}
\toprule
$T$ [MeV]	&	$\kappa_E/T^3$		&	$\kappa_B/T^3$		& $\kappa_c/T^3$		&	$2\pi T D_{s}^{c}$	&	$\kappa_b/T^3$		& $2\pi T D_{s}^{b}$ \\
\midrule
163	 & 	 $14.3^{+3.1}_{-3.3}$	 & 	 $8.0^{+4.4}_{-3.4}$	 & 	 $15.6^{+3.3}_{-3.7}$	 & 	 $1.0^{+0.3}_{-0.2}$	 & 	 $14.9^{+3.2}_{-3.5}$	 & 	 $0.9^{+0.3}_{-0.2}$	 \\
174	 & 	 $12.0^{+2.9}_{-2.9}$	 & 	 $7.7^{+3.3}_{-3.1}$	 & 	 $13.2^{+3.3}_{-3.4}$	 & 	 $1.2^{+0.4}_{-0.2}$	 & 	 $12.5^{+3.1}_{-3.1}$	 & 	 $1.1^{+0.4}_{-0.2}$	 \\
188	 & 	 $9.3^{+2.9}_{-2.9}$	 & 	 $4.9^{+2.2}_{-2.2}$	 & 	 $10.1^{+3.3}_{-3.2}$	 & 	 $1.6^{+0.8}_{-0.4}$	 & 	 $9.6^{+3.1}_{-3.0}$	 & 	 $1.5^{+0.7}_{-0.4}$	 \\
195	 & 	 $8.7^{+2.9}_{-2.8}$	 & 	 $4.7^{+2.1}_{-2.1}$	 & 	 $9.5^{+3.3}_{-3.2}$	 & 	 $1.7^{+0.9}_{-0.4}$	 & 	 $9.1^{+3.0}_{-3.0}$	 & 	 $1.5^{+0.8}_{-0.4}$	 \\
220	 & 	 $6.8^{+2.4}_{-2.6}$	 & 	 $4.8^{+2.2}_{-2.3}$	 & 	 $7.7^{+2.9}_{-3.0}$	 & 	 $2.2^{+1.4}_{-0.6}$	 & 	 $7.1^{+2.7}_{-2.8}$	 & 	 $2.0^{+1.2}_{-0.5}$	 \\
251	 & 	 $5.8^{+2.1}_{-2.7}$	 & 	 $3.3^{+1.6}_{-1.7}$	 & 	 $6.6^{+2.5}_{-3.1}$	 & 	 $2.8^{+2.4}_{-0.8}$	 & 	 $6.1^{+2.2}_{-2.9}$	 & 	 $2.3^{+2.1}_{-0.6}$	 \\
293	 & 	 $4.4^{+2.2}_{-2.2}$	 & 	 $2.3^{+1.2}_{-1.3}$	 & 	 $5.0^{+2.5}_{-2.6}$	 & 	 $3.6^{+4.9}_{-1.0}$	 & 	 $4.6^{+2.3}_{-2.3}$	 & 	 $3.1^{+3.1}_{-1.0}$	 \\
352	 & 	 $3.7^{+1.6}_{-2.0}$	 & 	 $1.8^{+1.1}_{-1.1}$	 & 	 $4.2^{+1.9}_{-2.4}$	 & 	 $4.9^{+6.4}_{-1.5}$	 & 	 $3.9^{+1.7}_{-2.2}$	 & 	 $3.8^{+4.8}_{-1.2}$	 \\
400	 & 	 $3.1^{+1.4}_{-1.8}$	 & 	 $1.9^{+1.2}_{-1.1}$	 & 	 $3.7^{+1.8}_{-2.2}$	 & 	 $6.0^{+8.5}_{-1.9}$	 & 	 $3.4^{+1.6}_{-2.0}$	 & 	 $4.5^{+6.4}_{-1.4}$	 \\
444	 & 	 $2.9^{+1.4}_{-1.7}$	 & 	 $1.5^{+1.0}_{-1.0}$	 & 	 $3.4^{+1.7}_{-2.1}$	 & 	 $6.9^{+10.2}_{-2.3}$	 & 	 $3.1^{+1.5}_{-1.9}$	 & 	 $5.0^{+7.5}_{-1.7}$	 \\
500	 & 	 $2.5^{+1.2}_{-1.6}$	 & 	 $1.2^{+0.9}_{-0.8}$	 & 	 $2.9^{+1.5}_{-1.8}$	 & 	 $8.6^{+13.8}_{-2.9}$	 & 	 $2.7^{+1.4}_{-1.7}$	 & 	 $6.0^{+9.8}_{-2.0}$	 \\
1000 & 	 $1.2^{+0.8}_{-0.8}$	 & 	 $0.7^{+0.7}_{-0.5}$	 & 	        ------       	 & 	                          ------ 	 & 	 $1.4^{+1.0}_{-0.9}$	 & 	 $13.9^{+24.8}_{-5.9}$	 \\
10000& 	 $0.5^{+0.3}_{-0.5}$	 & 	 $0.3^{+0.4}_{-0.2}$	 & 	    ------               & 	                          ------  	 & 	     ------              & 	          ------           \\
\bottomrule
\end{tabular}
\caption{Overall results and temperature dependence of the electric and magnetic momentum diffusion coefficients $\kappa_{E,B}$ (second and third columns), the momentum total diffusion coefficient $\kappa$ defined in Eq.~\eqref{eq:kappaE_B}, and the spatial diffusion coefficient $2\pi T D_{s}$ defined in Eq.~\eqref{eq:D_s}, for the charm quark (middle columns) and the bottom quark (right columns). The results are the mean values computed from the various models used to determine the mean squared thermal velocity and momentum (see text).}
\label{tab:kappa_D_final}
\end{table}

\begin{figure}[ht]
    \centerline{
        \includegraphics[width=0.4\textwidth]{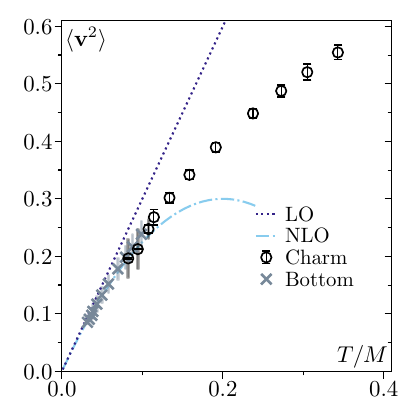}
        \includegraphics[width=0.4\textwidth]{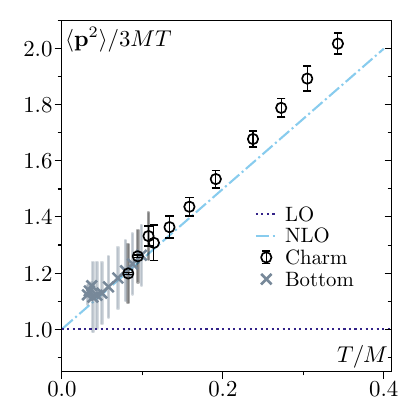}
    }
    \caption{The mean squared thermal velocity and the scaled mean squared thermal momentum as a function of $M/T$ for charm and bottom quarks using Model 1 (see text). We also show the expected temperature dependence at LO and NLO in the $T/M$ expansion by dotted and dashed-dotted lines, respectively. The different shades correspond to the systematic differences that arise from comparing Model 1 with Model 2 for charm quarks, and Model 1 with Models 2 and 3 for bottom quarks. 
    }
    \label{fig:v2_p2}
\end{figure}

As discussed in the main text, a quasi-particle description is necessary to estimate the heavy-quark diffusion coefficient beyond LO in the inverse heavy-quark mass. In this section of the appendix, we detail the quasi-particle model used to compute $\vv$ and $\pp$, based on available lattice QCD results. A model of this kind was first introduced in Ref.~\cite{Petreczky:2008px} to describe the temperature dependence of vector and axial-vector current susceptibilities for heavy (charm and bottom) quarks in quenched QCD. In Ref.~\cite{Petreczky:2008px} it was shown that $\vv$ obtained from the ratio of susceptibility of the vector current and the quark number susceptibility calculated on the lattice agrees with the quasi-particle model.

Generalized charm susceptibilities have been computed throughout the years in 2+1-flavor lattice QCD \cite{Bazavov:2014yba, Bazavov:2023xzm, Kaczmarek:2025dqt, Bellwied:2015lba}. Below the crossover temperature, their behavior can be understood within the framework of a hadron resonance gas (HRG) model. In this approach, the particle content includes not only experimentally observed charm hadrons listed by the Particle Data Group, but also additional states predicted by quark models \cite{Bazavov:2014yba,Bazavov:2023xzm,Kaczmarek:2025dqt}. This extended version is known as the QM-HRG model.

Above the chiral crossover temperature, charm susceptibilities can be described using a model in which the total charm pressure is given by the sum of partial pressures from charm mesons, charm baryons, and charm quarks \cite{Mukherjee:2015mxc,Bazavov:2023xzm,Kaczmarek:2025dqt}:
\be
    p_C(T)=p_C^M(T)+p_C^B(T)+p_C^q(T)\,.
    \label{eq:charm_pressure}
\ee
At sufficiently high temperatures, the partial pressures of charm mesons and baryons are expected to vanish, i.e., $p_C^M \simeq p_C^B \simeq 0$, while the charm-quark contribution $p_C^q$ approaches its ideal gas limit. Lattice QCD results confirm that $p_C^M(T)$ and $p_C^B(T)$ become small for $T>190$ MeV, and the total charm pressure is then dominated by the quark contribution \cite{Mukherjee:2015mxc,Bazavov:2023xzm,Kaczmarek:2025dqt}. Thus, at these temperatures, the generalized charm susceptibilities can be accurately described by a charm-quark gas.

In this study, we use continuum extrapolated lattice QCD results for the second and fourth order dimensionless charm susceptibilities as calculated in Refs.~\cite{Bellwied:2015lba} and \cite{Kaczmarek:2025dqt}, respectively. These susceptibilities are defined as:
\be
    \hat \chi_n^C=\frac{\partial^n (p_C(T)/T^4)}{\partial(\mu_C/T)^n},~n=2,4\,.
    \label{eq:susceptibilities_derivatives}
\ee
Here $\mu_C$ is the charm chemical potential. 
Due to the large charm-quark mass, we can use Boltzmann statistics, for which 
$\hat \chi_4^C=\hat \chi_2^C\equiv\chi_c/T^2=p_C(T)/T^4$. This relation is supported by lattice QCD results \cite{Bazavov:2014yba}.

We use interpolations to obtain values of the susceptibilities at the temperatures relevant to our analysis. Figure~\ref{fig:chi-charm} shows these interpolated values alongside the predictions from the QM-HRG. Specifically in the figure, we highlight the interpolated value of the charm susceptibility at $T=188$ MeV used in our study. While the QM-HRG model does not fully describe the lattice results for the generalized susceptibilities across all temperatures \cite{Kaczmarek:2025dqt}, it is noteworthy that the susceptibilities agree within errors with the QM-HRG predictions up to approximately $188$ MeV.

Since charm quarks dominate the relevant degrees of freedom for $T>190$ MeV, we use a simple quasi-particle model defined by Eqs.~\eqref{eq:qmp_chi}, \eqref{eq:qmp_vsq}, and \eqref{eq:qmp_psq}. In this regime, the quasi-particle mass $M_c(T)$ is determined directly from the lattice QCD results for the charm susceptibility $\hat \chi_2^C$, as shown in Fig.~\ref{fig:chi-charm} and listed in Tab.~\ref{tab:thermal_variables_charm}. Using these values of $M_c(T)$, we then compute the mean squared thermal velocity $\vv$ and mean squared thermal momentum $\pp$ for temperatures at and above $195$ MeV. The corresponding uncertainties are determined by the statistical errors on $\hat \chi_2^C$. This is the same approach used in the previous HotQCD study \cite{Altenkort:2023eav}.

For temperatures below $190$ MeV, we instead use the lattice QCD results for the charm-quark partial pressure $p_C^q(T)$ from Ref.~\cite{Kaczmarek:2025dqt} to determine the quasi-particle mass. This is done by using Eq.~\eqref{eq:qmp_chi} and replacing $\chi_c$ with $p_C^q/T^2$. In this low-temperature regime, charm meson and charm baryon contributions become relevant and must be included. For this purpose, we adopt the 1S1P-HRG model introduced in Ref.~\cite{Kaczmarek:2025dqt}, which includes the ground-state charm hadrons and their parity partners (first orbital excitations) \cite{Ebert:2011kk,Chen:2022asf}. This model has been shown to give a fair description of the lattice results for $p_C^M$ and $p_C^B$ \cite{Kaczmarek:2025dqt}. Using the charm-quark mass $M_c(T)$ extracted from $p_C^q(T)$ and the hadron masses listed in the 1S1P-HRG model, we calculate $\vv$ and $\pp$ for $T<190$ MeV. The results are summarized in Tab.~\ref{tab:thermal_variables_charm} and are referred to as Model 1, with corresponding quantities labeled using the subscript $c_1$. For sufficiently large masses, the relation $\pp=3 M_c T$ holds. Therefore, in the table we report the scaled quantity $\hat \pp=\pp/(3M_c T)$ to show the proximity of our results to the heavy-mass approximation. In fact, in order to determine $D_s$ from $\kappa$ we only need the value of $\pp/(3M_c T)$, cf. Eq.~\eqref{eq:D_s}. 

To assess the sensitivity of our results to the modeling assumptions for $T<190$ MeV, we also compute $\vv$ and $\hat \pp$ using a simplified quasi-particle model that includes only charm quarks. We refer to this as Model 2. The resulting charm-quark mass is somewhat smaller, as shown in Fig.~\ref{fig:chi-charm} and in Tab.~\ref{tab:thermal_variables_charm}. The corresponding values of $\vv$ and $\pp$ are also given in Tab.~\ref{tab:thermal_variables_charm}, labeled with the subscript $c_2$. A direct comparison between the results from Model 1 and Model 2 is shown in Fig.~\ref{fig:v2_p2_compare}. The differences are at most 10\%. We take half the difference as an estimate of the systematic uncertainty in the determination of $\vv$ and $\pp$ for $T<190$ MeV. This systematic error is subdominant compared to other sources of uncertainty in the extraction of $\kappa$ and $D_s$.

There are currently no lattice QCD results on bottom-quark susceptibilities in 2+1 flavor QCD to directly constrain the bottom quasi-particle mass $M_b(T)$. However, as discussed above for the charm quark, the charm susceptibility $\chi_c$ is well described by the QM-HRG model up to temperatures of $T=190$ MeV. Therefore, for $T<190$ MeV, it seems like a plausible choice to use the QM-HRG model to estimate the bottom susceptibility $\chi_b$.

To compute $\chi_b$ within QM-HRG, we include the known open-beauty hadrons listed in the Particle Data Group, and supplement the spectrum with additional states predicted by relativistic quark model calculations \cite{Ebert:2011kk,Chen:2022asf}, as done for the charm sector. Above the crossover temperature, we assume the same set of bottom degrees of freedom as for charm, and write the total bottom susceptibility as:
\be
    \chi_b(T)\cdot T^2=p_B(T)=p_B^M(T)+p_B^B(T)+p_B^q(T)\,,
    \label{eq:bottom_pressure}
\ee
where $p_B^M(T)$, $p_B^B(T)$ and $p_B^q(T)$ represent the partial pressures from bottom mesons, bottom baryons, and bottom quarks respectively. For the hadronic contributions $p_B^M(T)$ and $p_B^B(T)$, we use the 1S1P-HRG model, which again includes ground-state open-beauty hadrons and their first orbital excitations. For the spectrum of beauty hadrons we use the masses from Particle Data Group and relativistic quark model \cite{Ebert:2011kk,Chen:2022asf}. The corresponding pressures in QM-HRG and 1S1P-HRG are shown in Fig.~\ref{fig:preassures}.
With $\chi_b(T)$ from QM-HRG and $p_B^M(T)$ and $p_B^B(T)$ from 1S1P-HRG, we isolate the quark contribution $p_B^q(T)$, and use Eq.~\eqref{eq:qmp_chi} replacing $\chi_b(T)\rightarrow p_B^q(T)$ on the left hand side to extract the quasi-particle bottom-quark mass $M_b(T)$. The resulting values are listed in Table ~\ref{tab:thermal_variables_bottom}.

As in the charm sector, we assume that for $T>190$ MeV the hadronic contributions $p_B^M(T)$ and $p_B^B(T)$ are close to zero, and the degrees of freedom are dominated by bottom quarks. Furthermore, we assume that $M_b$ becomes temperature-independent above this threshold, as the QM-HRG model cannot be trusted at higher temperatures and we have no additional information to extract the temperature dependence of $M_b$. The corresponding values of $\vv$ and $\pp$ are given in Tab.~\ref{tab:thermal_variables_bottom}, and we refer to this as Model 1.

To test the sensitivity of $\vv$ and $\pp$ in the bottom sector to our model assumptions, we also consider two alternative models. In Model 2, we assume that bottom quarks are the only relevant degrees of freedom for $T<190$ MeV and extract the corresponding values of $M_b$ using Eq.~\eqref{eq:qmp_chi_ext}. As a result, the extracted bottom-quark masses are somewhat smaller than in Model 1. In Model 3, we assume that for $T>190$ MeV the bottom-quark mass is fixed at 4.5 GeV, independent of temperature. The corresponding values of $\vv$ and $\pp$ for Models 2 and 3 are provided in Tab.~\ref{tab:thermal_variables_bottom} within their respective temperature ranges. The differences relative to Model 1 are less than 10\% for $\vv$ and less than a few percent for $\hat \pp$. We use half the difference between Models 1 and 2 as an estimate of the systematic uncertainty in the determination of thermal variables for $T<190$ MeV, and the difference between Models 1 and 3 as the estimate of systematic uncertainty for $T\ge 195$ MeV.

We note that the mean squared thermal velocity for charm and bottom quarks can also be estimated using the NLO results for the massive vector correlation function \cite{Burnier:2012ze}, as discussed in Ref.~\cite{Laine_massive_VC}, when the temperature is sufficiently large. For the temperature range $T=251-352$ MeV, we find that the resulting estimates for $\vv_c$ and $\vv_b$ are lower by approximately 5–10\% and 15–17\%, respectively, compared to our quasi-particle model results.

In Fig.~\ref{fig:v2_p2}, we summarize the temperature dependence of $\vv$ and $\pp$ for charm and bottom degrees of freedom, along with their estimated uncertainties, as a function of $T/M$. The shades on the data represent the systematic differences that arise from comparing Models 2 and 3 to Model 1. At leading order in the $T/M$ expansion, we have the simple relations $\vv=3 T/M$ and $\pp=3 M T$. Our results show significant deviations from these LO predictions. However, including the NLO terms in the expansion in $T/M$ \cite{Bouttefeux:2020ycy} substantially improves the agreement, as can be seen in the figure.

In Tab.~\ref{tab:kappa_D_final} we present our final estimates of $\kappa_E$ and $\kappa_B$. Using these values, as well as the values of $\vv$ and $\pp$ obtained above, we compute the momentum and spatial heavy-quark diffusion coefficients for charm and bottom quarks, presented also in Tab.~\ref{tab:kappa_D_final}. The uncertainties in $\vv$ and $\pp$ are included in the total error estimates of $\kappa$ and $D_s$. These uncertainties, however, are significantly smaller than those associated with $\kappa_E$ and $\kappa_B$, as can also be seen in Tab.~\ref{tab:kappa_D_final}.

\acknowledgments
\label{sec:acknowledge}
This material is based on work supported by the U.S. Department of Energy, Office of Science, Office of Nuclear Physics under Contract No. DE-SC0012704 and by the Scientific Discovery through Advanced Computing (SciDAC) award "Fundamental Nuclear Physics at the Exascale and Beyond" and the Topical Collaboration in Nuclear Theory "Heavy-Flavor Theory (HEFTY) for QCD Matter".

Part of the analysis contained in this publication was based on the AnalysisToolBox code~\cite{AnalysisToolbox} developed by the HotQCD collaboration~\cite{Clarke:2023sfy}. We thank Luis Altenkort for providing his analysis code~\cite{luhuhis}. 

The authors acknowledge support from the Deutsche Forschungsgemeinschaft (DFG) through the CRC-TR 211 "Strong-interaction matter under extreme conditions" (Project No. 315477589 - TRR 211). J.H.W.'s research is also funded by the DFG under project number 417533893/GRK2575 "Rethinking Quantum Field Theory". J.H.W. acknowledges the support by the State of Hesse within the Research Cluster ELEMENTS (Project ID 500/10.006). R.L. was supported by the Ministry of Culture and Science of the State of Northrhine Westphalia (MKW NRW) under the funding code NW21-024-A (NRW-FAIR).

This research used computing time provided by the ALCC and INCITE programs at the Oak Ridge Leadership Computing Facility, a DOE Office of Science User Facility supported under contract no. DE-AC05-00OR22725; the National Energy Research Scientific Computing Center (NERSC), a DOE Office of Science User Facility at Lawrence Berkeley National Laboratory under Contract No. DE-AC02-05CH11231; PRACE awards on JUWELS at GCS@FZJ, Germany; Marconi100 at CINECA, Italy; and the LUMI-G supercomputer at the Finnish IT Center for Science (CSC). Computations were also performed on facilities of the USQCD Collaboration funded by the DOE Office of Science, in particular on the 21-g cluster at the Thomas Jefferson National Accelerator Facility (JLAB). In addition, parts of this work were performed on the GPU cluster of the University of Bielefeld, supported by HPC.NRW.

\textbf{Data Availability Statement}: The raw and derived data of this study are openly available in~\cite{bollweg2025data}. The gauge field configurations are available on request and will be openly available on the International Lattice Data Grid (ILDG) at a later stage.

\textbf{Code Availability Statement}: The lattice generation and GF analysis presented in this work were performed using SIMULATeQCD~\cite{HotQCD:2023ghu} and the MILC code. The analysis code is openly available in~\cite{AnalysisToolbox,luhuhis,bollweg2025data}.


\bibliographystyle{JHEP}

\providecommand{\href}[2]{#2}\begingroup\raggedright\endgroup

\end{document}